\documentclass[traditabstract]{aa} 
\usepackage{graphicx}
\usepackage{natbib}
\usepackage{amsmath}

\usepackage[applemac]{inputenc} 

\begin{document}
   \title{An optimized correlation function estimator for galaxy surveys}

   \author{M. Vargas-Maga\~na\inst{1} \and
		  J. E. Bautista\inst{1}\and
        		  J.-Ch. Hamilton\inst{1}\and
		  N.G. Busca\inst{1}\and
		  É. Aubourg\inst{1}\and
		  A. Labatie\inst{2}\and
		  J.-M.~Le~Goff\inst{3}	 \and 
		  S. Escoffier\inst{4} \and
		  M. Manera\inst{5} \and
		  C. K. McBride\inst{6} \and  
		  D. P. Schneider \inst{7}$^{,}$ \inst{8}
		  Ch. N. A. Willmer\inst{9}		  	  
          }

   \institute{APC, Astroparticule et Cosmologie, Université Paris Diderot, CNRS/IN2P3, CEA/Irfu, Observatoire de Paris, Sorbonne Paris Cité, 10, rue Alice Domon et Léonie Duquet, 75205 Paris Cedex 13, France
             \and
          Laboratoire AIM, CEA/DSM-CNRS-Université Paris Diderot, IRFU, SEDI-SAP, Service d'Astrophysique, Centre de Saclay, F-91191 Gif-Sur-Yvette cedex, France
	\and
             CEA centre de Saclay, irfu/SPP, F-91191 Gif-sur-Yvette, France
         \and 
	CPPM, Aix-Marseille Université, CNRS/IN2P3, Marseille, France
         \and
         	Institute of Cosmology and Gravitation, Portsmouth University, Dennis Sciama Building, Po1 3FX, Portsmouth, UK
	\and 
	Harvard-Smithsonian Center for Astrophysics, 60 Garden St., Cambridge, MA 02138, USA 
	\and
	Department of Astronomy and Astrophysics, The Pennsylvania State University, University Park, PA 16802, USA.
	\and
	Institute for Gravitation and the Cosmos, The Pennsylvania State University, University Park, PA 16802, USA.
	\and Steward Observatory, University of Arizona 933 N. Cherry Avenue Tucson, AZ, 85721, USA.
	}

   \date{Received xxxx / accepted xxxx}

  \abstract{Measuring the two-point correlation function of the galaxies in the Universe gives access to the underlying dark matter distribution, which is related to cosmological parameters and to the physics of the primordial Universe.    The estimation of the correlation function for current galaxy surveys makes use of the Landy-Szalay estimator, which is supposed to reach minimal variance. This is only true, however, for a vanishing correlation function.  We study the Landy-Szalay estimator when these conditions are not fulfilled and  propose a new estimator that provides the smallest variance for a given survey geometry. Our estimator is a linear combination of ratios between paircounts of data and/or random catalogues (DD, RR and DR). The optimal combination for a given geometry is determined by using lognormal mock catalogues. The resulting estimator is biased in a model-dependent way, but we propose a simple iterative procedure for obtaining an unbiased model- independent estimator.  
  Our method can be easily applied to any dataset and requires few extra mock catalogues compared to the standard Landy-Szalay analysis.
    Using various sets of simulated data (lognormal, second-order LPT and N-Body), we obtain a 20-25\% gain on the error bars on the two-point correlation function for the SDSS geometry and $\Lambda$CDM correlation function. When applied to SDSS data (DR7 and DR9), we achieve a similar gain on the correlation functions, which translates into a 10-15\% improvement over the estimation of the densities of matter $\Omega_m$ and dark energy $\Omega_\Lambda$ in an open $\Lambda$CDM model. The constraints derived from DR7 data with our estimator are similar to those obtained with the DR9 data and the Landy-Szalay estimator, which covers a volume twice as large and has a density that is three times higher. 
  }
    
   \keywords{Cosmology --
                Large Scale Structure --
                Baryonic Acoustic Oscillations
               }

\authorrunning{Vargas et al}

\maketitle

\section{Introduction}

 The distribution of galaxies in the Universe is an extremely rich source of information for cosmology. Indeed, galaxies 
trace the underlying dark matter distribution in a way that is typically described with a multiplicative factor known as the \emph{bias}. To a good approximation, this bias can be considered independent of scale.
On larger scales where fluctuations are haloesstill small, one can apply linear theory and have a direct access to cosmological parameters. On smaller scales, gravity acts in a non-linear manner and the galaxy clustering allows one to investigate the structuration of dark matter into haloes.
Observing the large-scale structure of the Universe is a promising approach for improving our understanding of its accelerated expansion  
observed by various cosmological probes in the past decade. The cosmic acceleration was initially proposed as a way to reconcile the apparent low matter content of the Universe with a flat geometry in a standard cold dark matter scenario~[\cite{efstathiou90}]. The first convincing measurement of cosmic acceleration came from observations that type Ia supernovae appeared less luminous than expected in a decelerating Universe~[\cite{riess98, perlmutter99}]. These observations can be accommodated by modifying general relativity on cosmological scales or, within a Friedmann-Lema\^itre-Robertson-Walker (FLRW) cosmology, by adding a {\em dark energy} component with a density $\Omega_X\sim 0.7$, a negative pressure, and a possibly evolving equation of state. Since then, the cosmic acceleration has  
been confirmed by other probes, including the cosmic microwave background (CMB) fluctuations~[\cite{WMAP, ACTlensing}], integrated Sachs–Wolfe (ISW) effect~[\cite{ISW}] and baryonic acoustic oscillations (BAO) (\cite{weinberg} for a general review and \cite{anderson} for the latest measurement). These data point towards a dark energy with a constant equation-of-state parameter, $w=-1$, or equivalently a pure cosmological constant.  BAO measurements are based on the observation of an acoustic peak in the correlation function of the matter density fluctuations, corresponding to the acoustic horizon at the epoch of matter-radiation decoupling~[\cite{EisensteinHu}]. The acoustic scale is used as a standard ruler at various redshifts, allowing for the measurement of the angular distance in the transverse directions and the expansion rate in the radial direction [\cite{reid2012}]. 

When investigating the large-scale structure of the Universe using galaxies, one needs large field-of-view deep galaxy surveys such as the Sloan Digital Sky Survey (SDSS-III) Baryon Oscillation Spectroscopic Survey (BOSS) ~[\cite{boss}],  
i.e. high density galaxy catalogues, where the radial positions of galaxies are measured by their redshifts. The two-point correlation function is commonly used for characterizing the large-scale structure within such galaxy surveys. One does not directly measure the density within the survey volume, but samples this density through galaxy locations, makes the estimation of the two-point correlation function more complex. 
The observed quantity is the average number of neighbours at a given distance in the survey volume and is biased by the fact that galaxies near the edges of the catalogue volume have less neighbours than they should have, which needs to be corrected for in an optimal way. This issue does not occur, for example when directly measuring a function of the matter density through the Lyman-$\alpha$ forest of distant quasars~[\cite{slosar}]. 

\begin{figure*}[t]
   \centering
   \resizebox{\hsize}{!}{\includegraphics{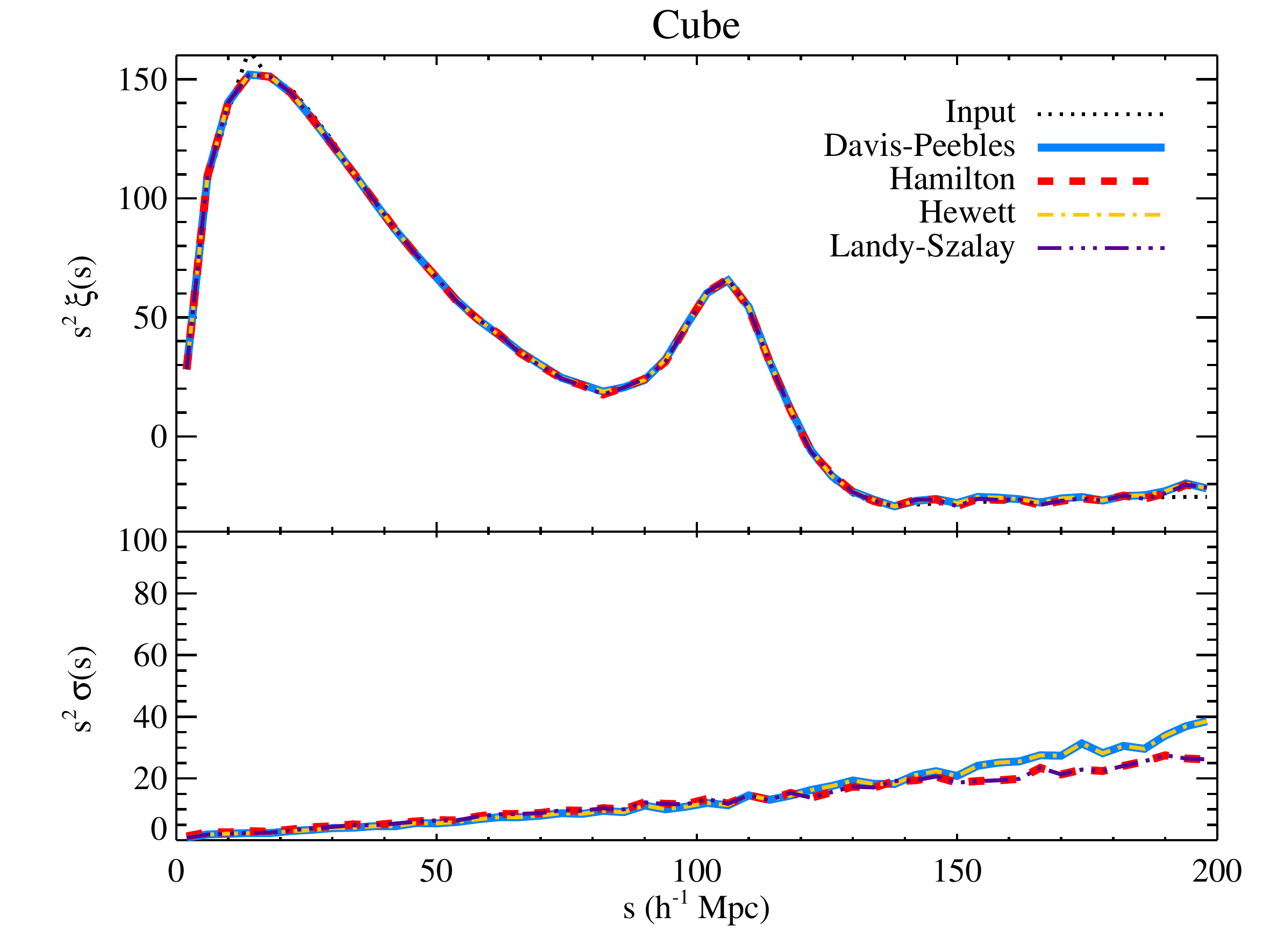}~\includegraphics{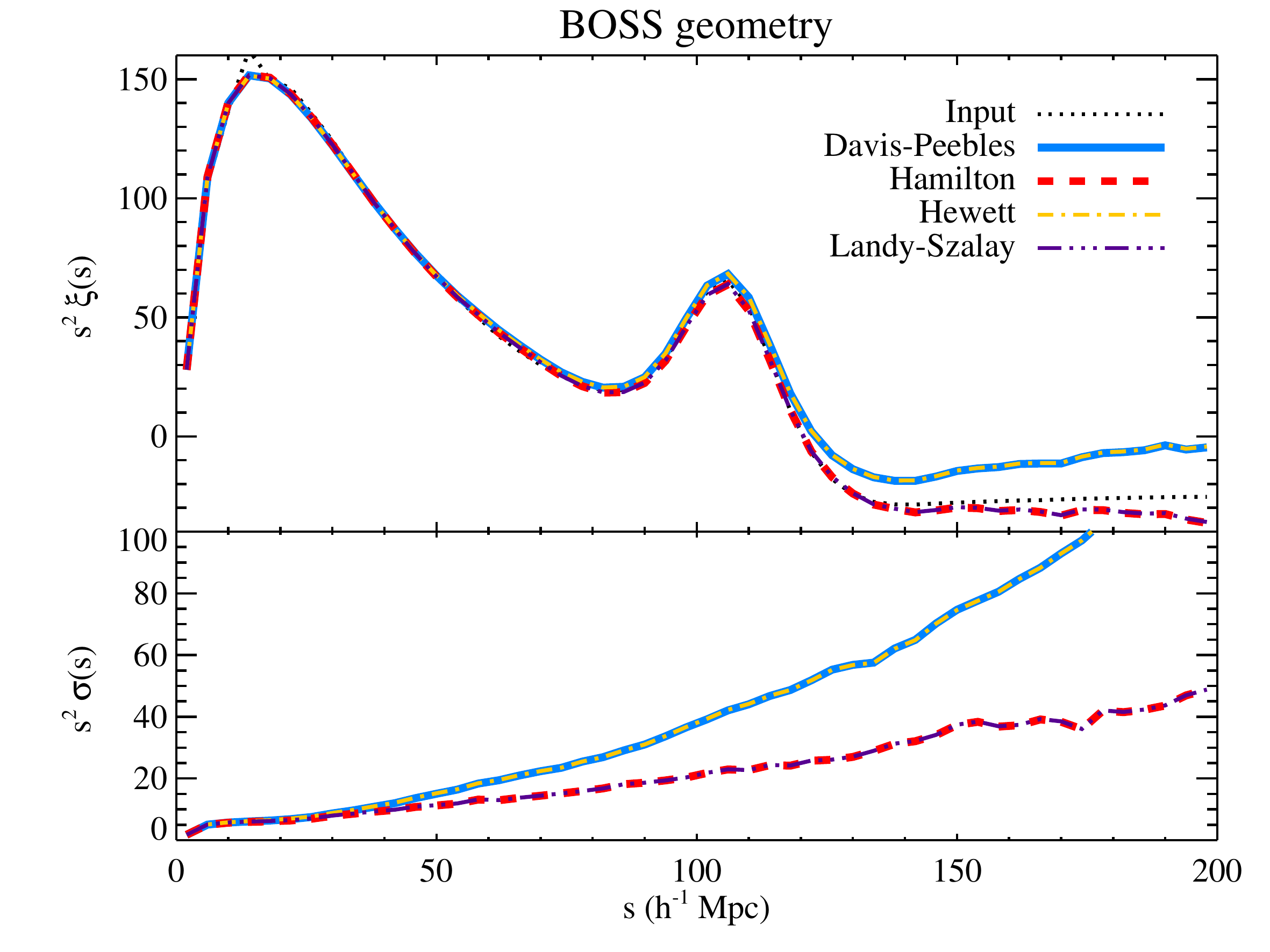}}
   \caption{{\bf Input (dotted line) and reconstructed (various colours and linestyles, see legend) two-point correlation function obtained using the various estimators available in the literature for a cubic geometry (left) or a realistic (BOSS DR9) survey volume (right). The bottom panels show the root mean square (RMS) of each estimator with corresponding colour and linestyle. In each case, the Hamilton and Landy-Szalay lines are exactly superposed as well as the Davis-Peebles and Hewett lines. (Coloured version of the figure available online).}
 }
              \label{FigBiasCube}%
    \end{figure*}

In this article, we introduce a novel estimator for the two-point correlation function of galaxies. Its performance can be optimized for a given galaxy survey geometry. In section~{\ref{motivations} we justify this effort, showing that various well-known estimators for the two-point correlation function have a residual bias and a variance that strongly depend on the survey geometry. The commonly used Landy-Szalay estimator~[\cite{landyszalay}] has been shown to have no bias or have minimal variance in the limit of a vanishing correlation function. We show that in realistic cases, where the correlation function is not zero, the Landy-Szalay estimator does not reach the Poisson noise limit.
For pedagogical reasons, we start in section~\ref{optimalestimator} with a simpler but biased version of our optimal estimator, and in section~\ref{sec:iterative} we develop a simple iterative procedure that allows the final estimator to be model-independent, with an improvement in the accuracy around 20-25\% with respect to the Landy-Szalay estimator. In section~\ref{data} we apply our final estimator to data from the SDSS-II Seventh Data Release (DR7) Luminous Red Galaxy sample and on the SDSS-III/BOSS DR9 ``CMASS " sample and show the improvement in the two-point correlation function measurement and cosmological parameters over previous analyses.


\section{Motivations for an optimized two-point correlation function estimator\label{motivations}}

\subsection{Commonly used estimators}
Estimators of the two-point correlation function $\xi(s)$ ($s$ being the comoving separation) have been studied by various authors [\cite{peebleshauser}, \cite{davispeebles}, \cite{hewett},\cite{hamilton},\cite{landyszalay}]. 
Generically, pair counts in data are compared to pair counts in random samples that follow the geometry of the survey. We assume a catalogue of $n_d$ objects in the data sample and $n_r$ in the random sample and then calculate three sets of numbers of pairs as a function of the binned comoving separation $s$\footnote{The number of pairs can be spherically averaged in the simplest approach. Its dependence on the angle with respect to the line of sight can be considered in a more elaborated analysis, in order to account for the sensitivity to angular distance in the transverse direction and $H(z)$ in the radial one (see~[\cite{cabre}] for details).}
\begin{itemize}
\item within the data sample, $dd(s)$ that can be normalized by the total number of pairs as\\
 $ DD(s) = \displaystyle\frac{dd(s)}{n_d(n_d-1)/2} $;
\item within the random sample, leading to $rr(s)$ normalized as\\
$RR(s) = \displaystyle\frac{rr(s)}{n_r(n_r-1)/2}$;
\item among both samples (cross correlation) leading to $dr(s)$ normalized as\\
 $DR(s) = \displaystyle\frac{dr(s)}{n_r n_d}$.
\end{itemize}

The most common estimators discussed in the literature are:
\begin{itemize}
\item $\hat{\xi}_{PH}(s)=\displaystyle\frac{DD}{RR}-1$\hfill[\cite{peebleshauser}]
\item $\hat{\xi}_{Hew}(s)=\displaystyle\frac{DD-DR}{RR}$\hfill[\cite{hewett}] 
\item $\hat{\xi}_{DP}(s)=\displaystyle\frac{DD}{DR}-1$\hfill[\cite{davispeebles}]
\item $\hat{\xi}_H(s)=\displaystyle\frac{DD\times RR}{DR^2}-1$\hfill[\cite{hamilton}] 
\item $\hat{\xi}_{LS}(s)=\displaystyle\frac{DD-2DR+RR}{RR}$\hfill[\cite{landyszalay}].
\end{itemize}

Some studies have compared the behaviour of the different two-point correlation function
estimators, mainly in the small-scale regime and using smaller samples. In \cite{ponsborderia},  six estimators were analyzed,  including both the Hamilton and
Landy-Szalay estimators, and the authors did not find any outstanding winner among those estimators. In \cite{kerscher1} and \cite{kerscher2},  nine estimators
were considered, and the estimators presenting the best properties were the Landy-Szalay and Hamilton
estimators.

\subsection{Relative performances of the common estimators}

To compare the performances of these estimators, we used two sets of 120 mock ca\-ta\-lo\-gues obtained from lognormal~[\cite{colesjones}] density field simulations containing about 271,000 galaxies in both a  cube of 1 $h^{-1}$ Gpc size and a  far more complex geometry corresponding to the BOSS (DR9) survey~[\cite{anderson}] which, roughly contains the same volume as the cube. In addition we used random catalogues with three time as many galaxies as the mock catalogues for both geometries. The cosmology used for the lognormal fields is taken from the Wilkinson Microwave Anisotropy Probe (WMAP) 7 years analysis~[\cite{WMAP}]. 

Figure ~\ref{FigBiasCube} shows the correlation function obtained with the different estimators for the cubic and DR9 geometries. We clearly see differences between the performances of the estimators in the cube and in the DR9, both in their mean result and in their root-mean-square errors, RMS (bottom panels with corresponding colour and linestyles). 
In the case of the DR9 geometry, the mean results obtained with the Peebles-Hauser, Davis-Peebles and Hewett estimators are more biased than the theory on large scales. 
Landy-Szalay and Hamilton estimators are much less biased than the others in this more complex geometry. Examining the RMS, all estimators have their accuracy degraded by the effects of geometry. Landy-Szalay and Hamilton again show best performances with the lowest variances in both geometries as expected.

\begin{figure*}[!t]
   \centering
   \resizebox{\hsize}{!}{\includegraphics{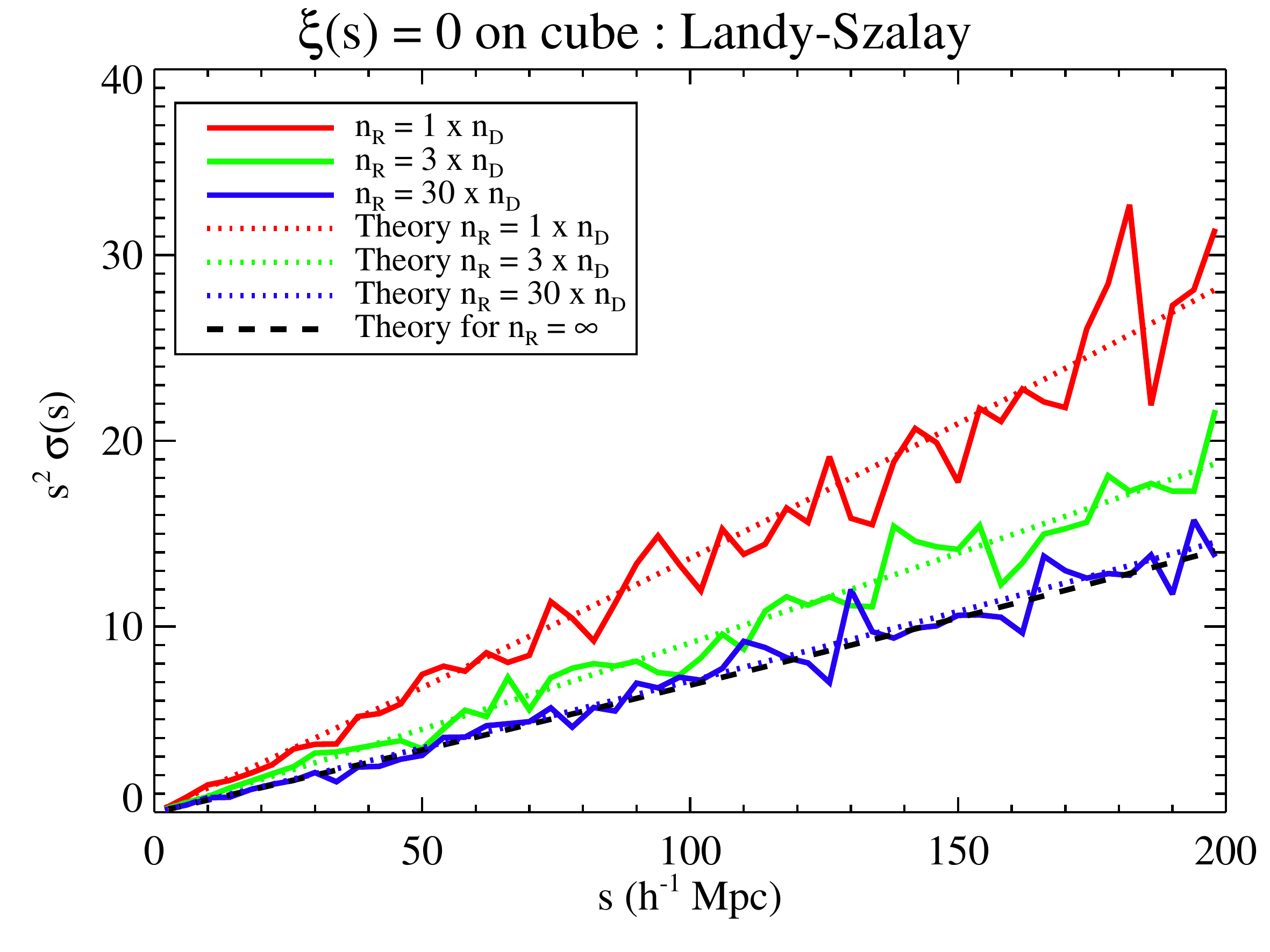}~\includegraphics{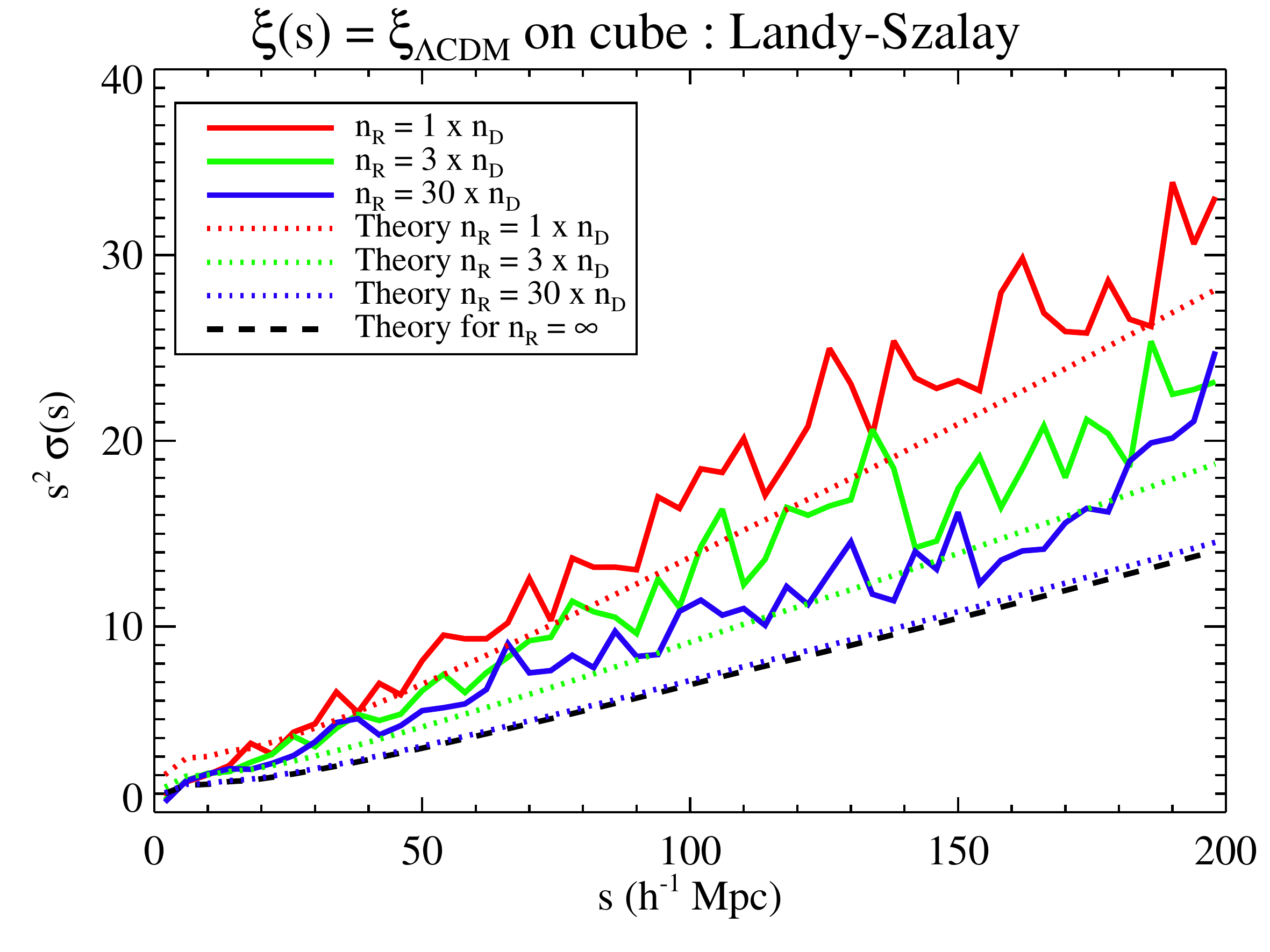}}
\caption{ RMS of the Landy-Szalay estimator for lognormal simulations in a cubic geometry and for a zero two-point correlation function (left) and for a $\Lambda CDM$ model (right). The upper, middle, and lower solid lines correspond to a random sample with 1, 3, and 30 times more galaxies, respectively than the data sample. Dotted lines show the Poisson noise associated with each number of randoms (in the same order as solid lines). The thick dashed line shows the limit corresponding to an infinite number of random galaxies. 
 (Coloured version of the figure available online)}
              \label{testLS}%
    \end{figure*}

\subsection{Optimality of Landy-Szalay estimator}
In the limit of an infinitely large random catalogue, for which the volume is much larger than the observed scales, 
and of a vanishing two-point correlation function (uniform galaxy distribution), the Landy-Szalay estimator is known to have no bias or have minimal variance. 
It is therefore used most widely in modern galaxy surveys [e.g., \cite{Eisenstein05}, \cite{percival},\cite{kazin2010},\cite{blake2},  \cite{anderson}, \cite{sanchez}]. 
In practice the volume of modern surveys is sufficiently large, and one can also produce a large enough random catalogue, 
but the correlation function to be measured is non-zero, so it is crucial to check residual bias and variance of estimators in the case of realistic non-zero correlation functions.

Using additional lognormal simulations, we investigated the RMS of the Landy-Szalay estimator as a function of the size of the random catalogue for both a zero correlation function and the one expected from the  Lambda Cold Dark Matter ($\Lambda$CDM) scenario. 
Fifty realizations were produced in both cases where a cubic geometry was used in order to be insensitive to the degradation due to the survey geometry. The resulting RMS are shown in Fig.~\ref{testLS}, along with the expectations for an optimal estimator (from equation 48 in ~\cite{landyszalay} 
accounting for the finite size of the random catalogue. It appears that, when the correlation function is not vanishing, the Landy-Szalay estimator does not reach the Poisson noise limit. 
This suggests that a better estimator can be found in the case of a non-vanishing correlation function and a more complicated survey geometry.

\section{An optimized estimator\label{optimalestimator}}
\subsection{General form and optimization criterion}

Our search for a better estimator started from the observation that the commonly used estimators are linear combinations of ratios of pair counts, $\vec{DD}$, $\vec{DR}$, and $\vec{RR}$ (hereafter the $s$ dependence is described by vectors), with the exception of the Hamilton estimator, that involves ratios of second-order products of pair counts. We therefore investigate an estimator which would be 
an optimal linear combination of all possible ratios $\vec{R}_i$ up to the second order. Table~\ref{tab:ratios} summarizes the six ratios at first order and the twelve at second order. The generic optimal estimator can then be expressed as

\begin{equation}
\hat{\vec{\xi}}^{opt}(\vec{c}) = c_0 + \sum^6_{i = 1} c_i \vec{R}_i^{(1)} + \sum^{18}_{i = 7} c_i \vec{R}_i^{(2)}	\;.
\label{eq:opt_est_definition}
\end{equation}

\begin{table}
\caption{The nineteen ratios formed by using pair counts up to second order.}
\label{tab:ratios}
\begin{center}
\begin{tabular}{ccc}
\hline
\hline
\multicolumn{3}{c}{0th order} \\
\multicolumn{3}{c}{1} \\
\hline
\multicolumn{3}{c}{1st-order terms $R^{(1)}$ } \\
$\displaystyle\frac{DD}{RR}$ & $\displaystyle\frac{DR}{RR}$ & $\displaystyle\frac{DR}{DD}$ \\ ~\\
$\displaystyle\frac{RR}{DD}$ & $\displaystyle\frac{RR}{DR} $& $\displaystyle\frac{DD}{DR}$ \\ ~\\
\hline
\multicolumn{3}{c}{ 2nd order terms $R^{(2)}$} \\
$\displaystyle\frac{DR \times RR}{DD^2}$ & $\displaystyle\frac{RR^2}{DD^2}$ & $\displaystyle\frac{DR\times DD}{RR^2}$ \\~\\
 $\displaystyle\frac{DD^2}{RR^2}$ & $\displaystyle\frac{DR^2}{RR^2}$ &  $\displaystyle\frac{DD^2}{DR^2}$ \\~\\
$\displaystyle\frac{RR^2}{DR^2}$ & $\displaystyle\frac{DD \times RR}{DR^2}$ & $\displaystyle\frac{RR^2}{DD \times DR}$ \\~\\
 $\displaystyle\frac{DR^2}{DD \times RR}$ & $\displaystyle\frac{DD^2}{DR \times RR}$ & $\displaystyle\frac{DR^2}{DD^2}$ \\~ \\
\hline
\hline
\end{tabular}
\end{center}
\end{table}

The nineteen $c_i$ coefficients are optimized lognormal to minimize the variance of the estimator for a given geometry. This optimization is done through a $\chi^2$ minimization using a large set of mock catalogues generated using lognormal fields, for which $DD$, $DR$, and $RR$ are stored, so that all the $R_i$ terms can be calculated. The $\chi^2$ is minimized with respect to the vector of parameters $\vec{c}$ as

\begin{equation}
\chi^2 = \sum_j \left[ \hat{\vec{\xi}}^\mathrm{opt}_j(\vec{c}) -\vec{\xi}_\mathrm{th}\right]^T \cdot N_\mathrm{LS}^{-1} \cdot \left[ \hat{\vec{\xi}}^\mathrm{opt}_j(\vec{c}) -\vec{\xi}_\mathrm{th}\right] \; ,
\label{eq:chi2}
\end{equation}

\noindent where the $j$ index stands for the $j$-th realization, $\hat{\vec{\xi}}^\mathrm{opt}$ is the vector of the values of the estimator in the comoving distance $s$ bins,and $\vec{\xi}_\mathrm{th}$ the vector for the theoretical input correlation function.
The quantity $N_\mathrm{LS}$ is the covariance matrix of fluctuations of $\vec{\xi}_j^\mathrm{LS}$ around the mean Landy-Szalay correlation function $\left< \vec{\xi}^\mathrm{LS}\right>$, the mean taken over the mock realizations. 

This approach will result in an estimator with a variance that is at most as large as that of the Landy-Szalay estimator, but which might have a significant bias. This bias with respect to the ``true'' correlation function will be referred to as {\em residual bias} in this article to distinguish it from the luminous matter bias with respect to dark matter.
The residual bias can be calculated and corrected for to an arbitrary precision 
for a given input correlation function (therefore for a given cosmological model).  However, this residual bias correction is model dependent and would only work perfectly when the input cosmology in the mock data matches the one to be measured in the real data. In section \ref{sec:iterative} we propose a simple iterative method that allows efficient circumvention of this problem with modest extra CPU time.

\subsection{Performances on simulations}
Using lognormal simulations, we produced 120 realizations of galaxy catalogues with a geometry similar to that of the SDSS-III/BOSS (DR9) survey [\cite{boss, anderson}]. 
The fiducial cosmology was defined by $h=0.7$, $\Omega_m=0.27$, $\Omega_\Lambda=0.73$, $\Omega_b=0.045$, $\sigma_8=0.8$, and $n_s=1.0$.
For each realization, we generated a random catalogue with the same geometry and calculated $\vec{DD}$, $\vec{DR}$, and $\vec{RR}$ for comoving separations between 0 and 200 $h^{-1}~\mathrm{Mpc}$ with bins of 4 $h^{-1}~\mathrm{Mpc}$. We then calculated the Landy-Szalay estimator for each simulation $\vec{\xi}_j^\mathrm{LS}$, the average estimator $\left< \vec{\xi}^\mathrm{LS}\right>$, and its covariance matrix $N_\mathrm{LS}$, empirically, i.e., from the dispersion of the individual realizations.

We then had all the ingredients required to minimize the $\chi^2$ in Eq.~\ref{eq:chi2} and obtain an  optimal estimator, which was done by limiting the $\chi^2$ to the region [40, 200] $h^{-1}~\mathrm{Mpc}$.
 This range corresponds to the typical interval used in BAO analyses. However, this choice is not fundamental, so our method could be applied to any other range depending on the purpose of the analysis.
The actual values of the coefficients $c_i$ are not particularly meaningful for two reasons: they depend on the geometry of the survey and are therefore not ``general''; in addition, the parameters are degenerate because the nineteen $\vec{R}_i$ terms are not independent. \\

 \begin{figure}
   \centering
   \resizebox{\hsize}{!}{\includegraphics{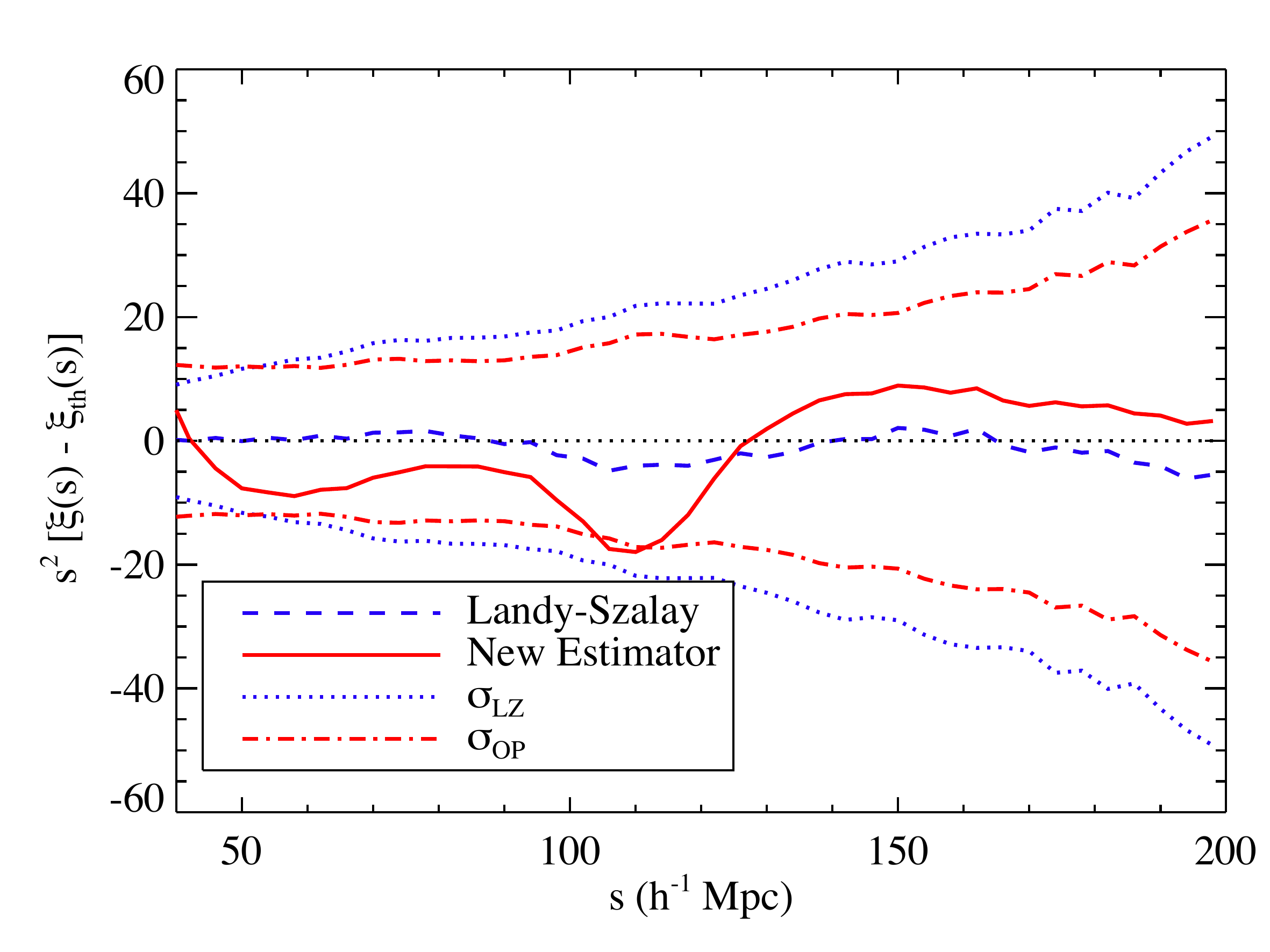}}
\caption{Residuals with respect to the input correlation function for the Landy-Szalay estimator (dashed-blue) and for the minimum variance estimator (solid-red). We also show the corresponding $\pm 1\sigma$ (RMS from lognormal simulations) for Landy-Szalay (dotted-blue) and for the minimum variance estimator (dot-dashed-red). Residuals and RMS are rescaled by {\bf $s^2$}. The optimal estimator has been obtained limiting the $\chi^2$ to the region [40, 200] $h^{-1}~\mathrm{Mpc}$. This range has been chosen accordingly to the typical range used in BAO analysis.
   (Coloured version of the figure available online).}
              \label{fig:estimator_bias}%
    \end{figure}
    
Figure ~\ref{fig:estimator_bias} compares our estimator to the Landy-Szalay estimator. The residual bias with respect to the theoretical input correlation and the RMS are shown for both estimators. The residual bias is defined as
\begin{equation}
\vec{B} = \left\langle \hat{\vec{\xi}}- \vec{\xi}_\mathrm{theory} \right\rangle.
\end{equation}  
These RMS are just the square roots of the diagonal elements of the estimator covariance matrix, calculated empirically from the individual lognormal realizations. The covariance matrix is defined as
\begin{equation}
C = \left\langle \left[ \vec{\xi} - \bar{\vec{\xi}} \right] \left[ \vec{\xi} - \bar{\vec{\xi}}\right]^T \right\rangle.
\end{equation}
The residual and the RMS are both rescaled by $s^2$.
The Landy-Szalay estimator essentially has no residual bias, while a significant residual bias is observed for our estimator, which, however, remains within the 1$\sigma$ range. 
In contrast, our optimized estimator appears to have smaller variances than the Landy-Szalay estimator in the region [40, 200] $h^{-1}~\mathrm{Mpc}$ , where the fit was performed.
Figure ~\ref{fig:covmat} shows the covariance and correlation matrices for the Landy-Szalay and optimized estimators. The latter have a smaller covariance matrix and no extra correlation between the bins.  The correlation matrix is defined as $c_{ij} =   C_{ij}/ \sqrt{C_{ii}} \sqrt{C_{jj}}$.

\begin{figure}
   \centering
   \includegraphics[width=0.49\columnwidth]{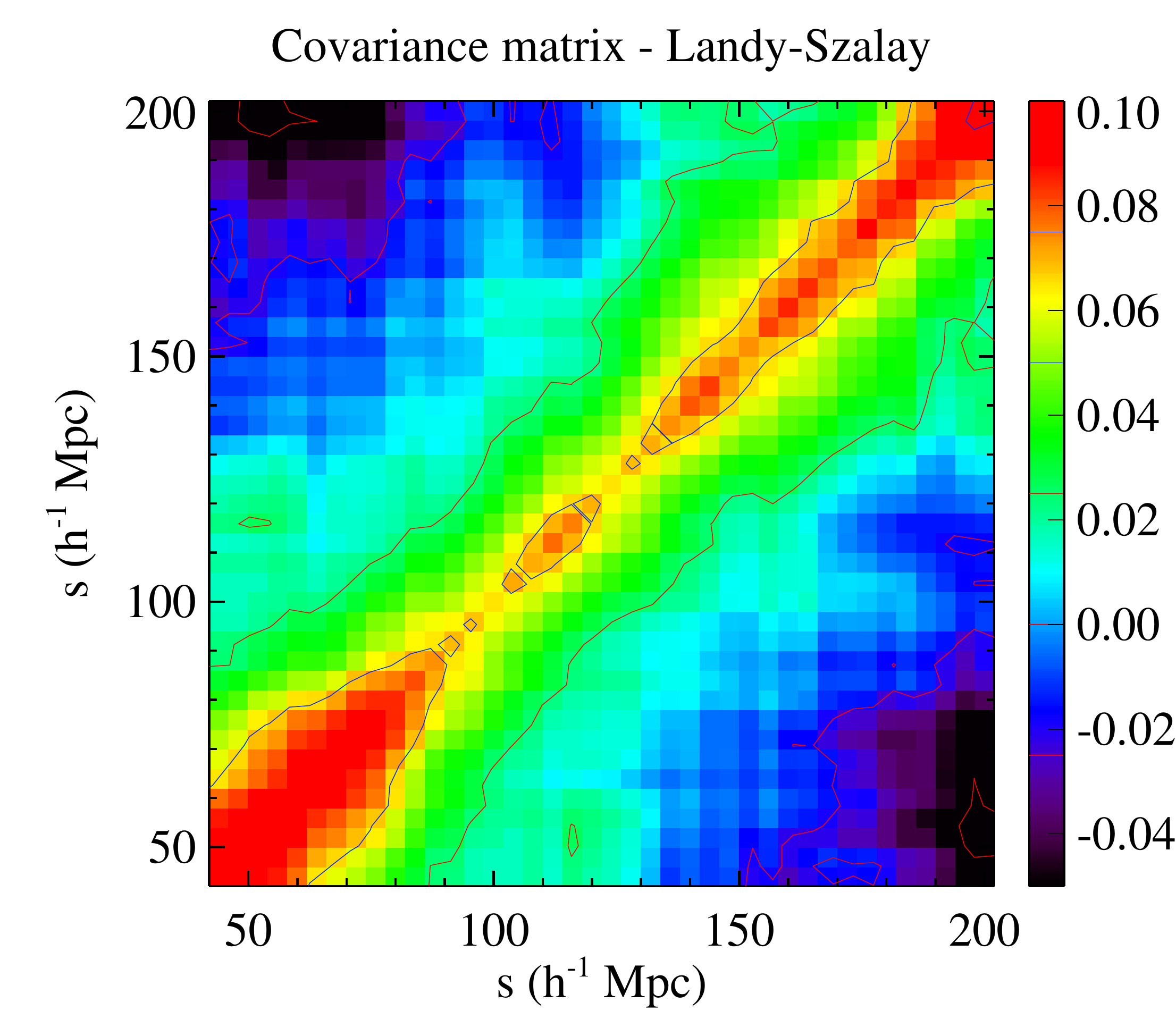}
   \includegraphics[width=0.49\columnwidth]{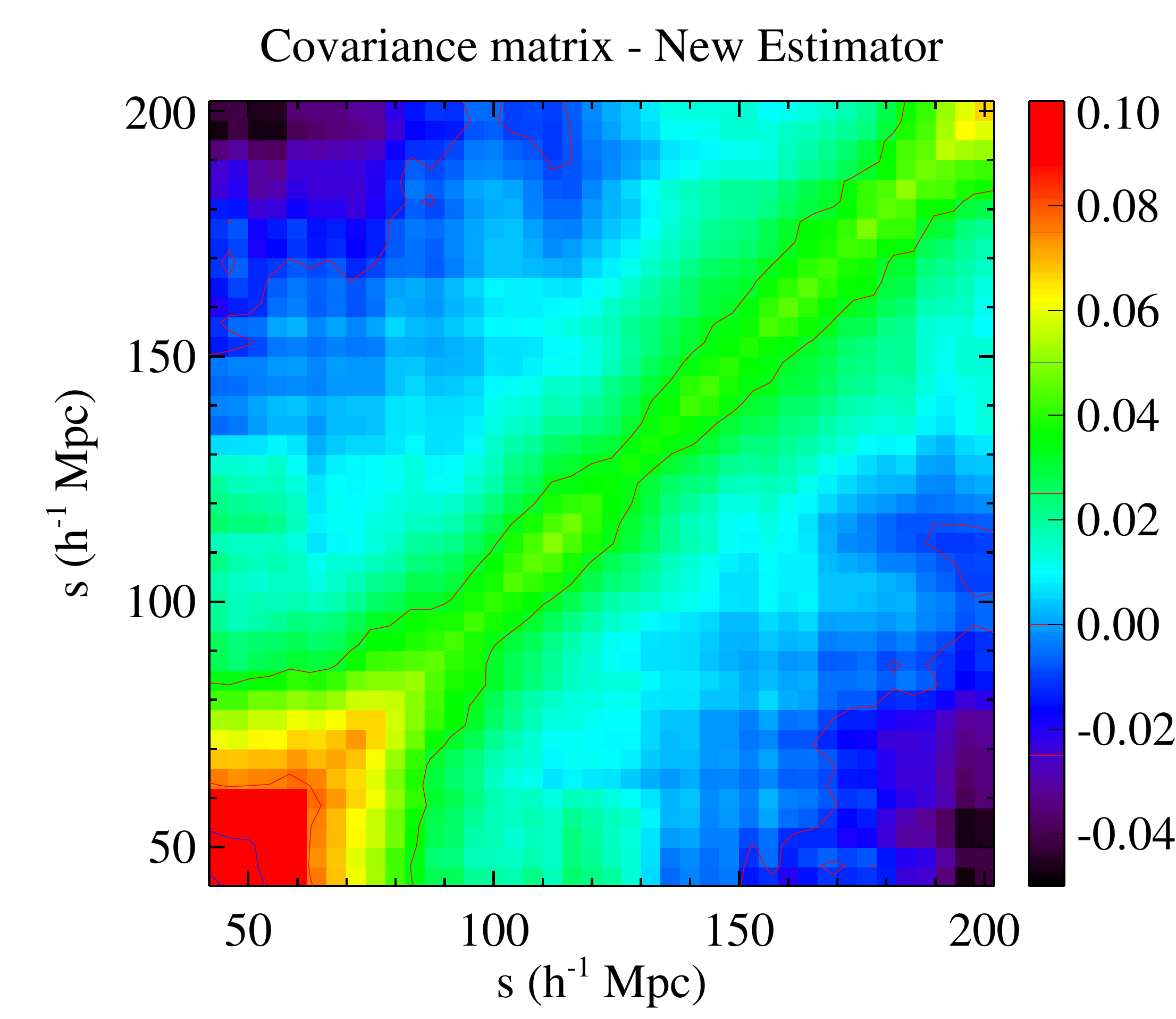}
   \includegraphics[width=0.49\columnwidth]{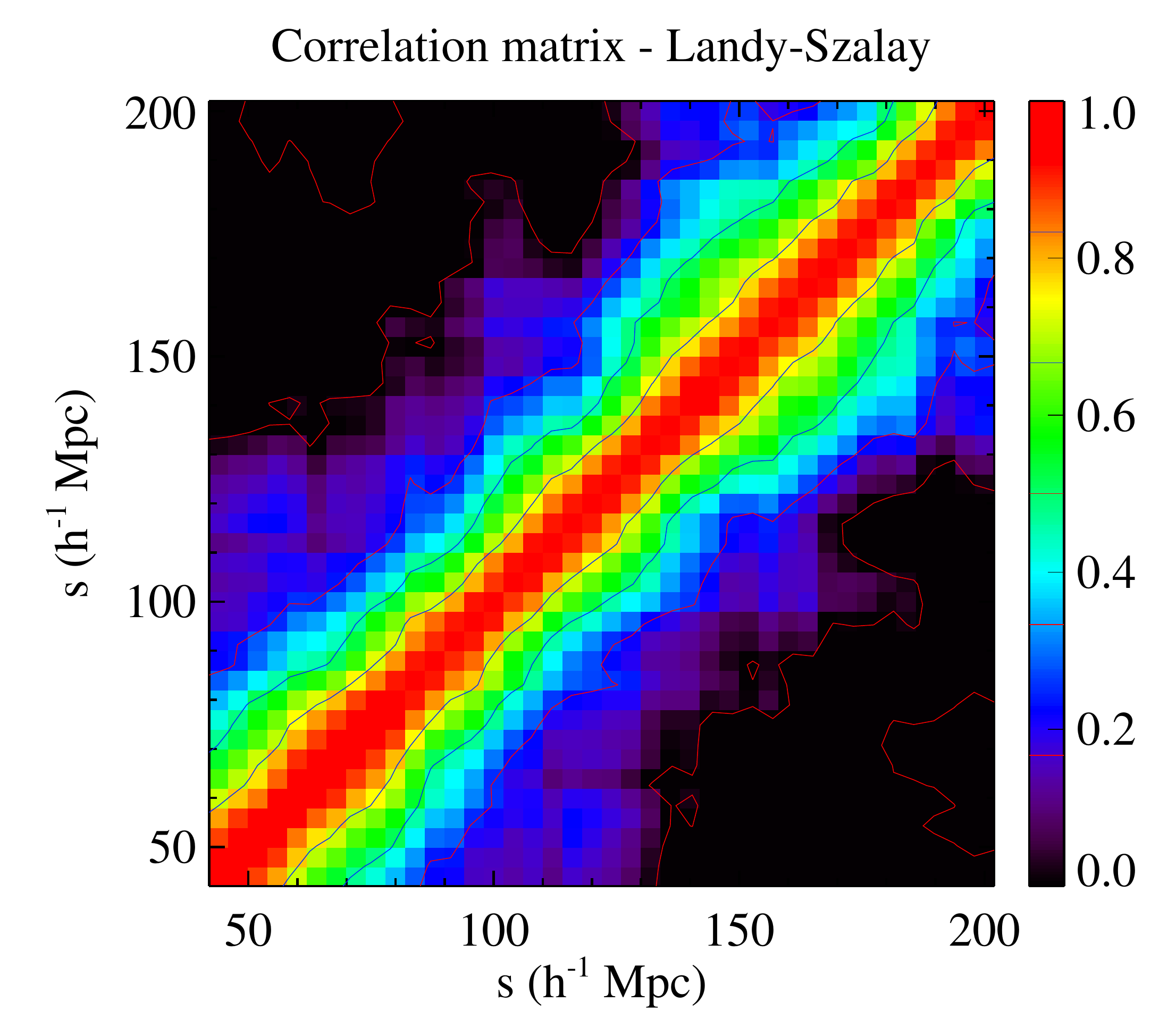}
   \includegraphics[width=0.49\columnwidth]{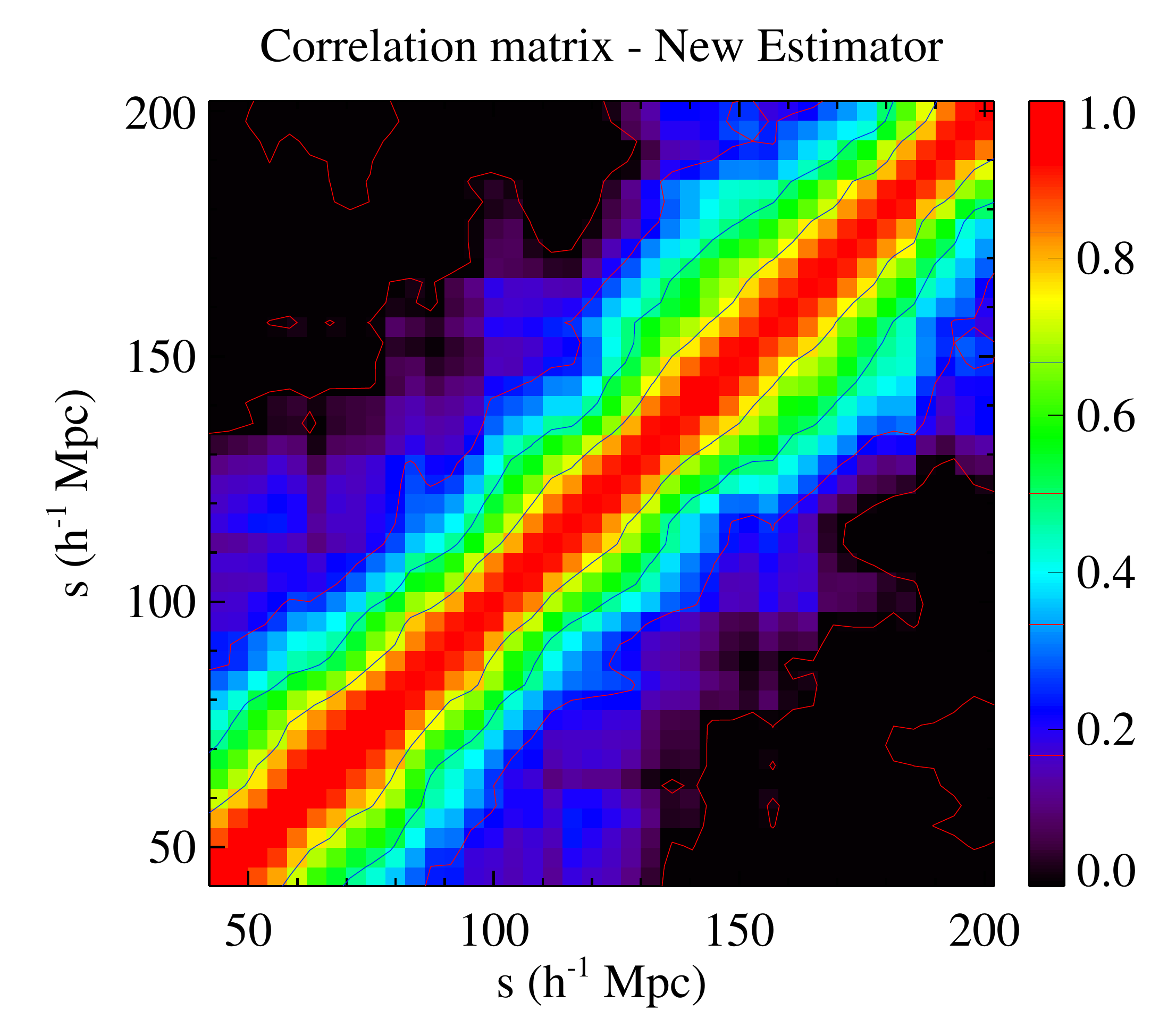}
   \caption{[Top panels] Covariance matrices multiplied by the square of the comoving distance. [Bottom panels] Correlation matrices for both the Landy-Szalay estimator (left panels) and the optimized estimator (right panels). 
   (Coloured version of the figure available online)
   }
              \label{fig:covmat}%
\end{figure}

\subsection{Model dependance}
Figures ~\ref{fig:estimator_bias} and ~\ref{fig:covmat} show that by correcting the optimized estimator by its average residual bias, which can be known with excellent accuracy by having a large number of mock realizations, one can achieve better accuracy on the correlation function than the Landy-Szalay estimator. 
Unfortunately, the residual bias exhibits a peak at the location of the BAO scale and therefore will be different in another cosmology: it is strongly model dependent. 
If one uses an estimator that is optimized with a set of simulations that assumed a cosmology different from the actual one, the peak position in the residual bias will be different from the one in the data, resulting in a strong distortion of the peak shape after bias correction and in a shift in its location. This is illustrated in Fig.~\ref{fig:res_alphabias}, and  discussed further. 
Fortunately, one can eliminate the cosmology dependence of the fitting, as described in the next section. 

\section{Iterative optimized estimator\label{sec:iterative}}
To transform the optimized estimator into a model-independent one, we investigated the possibility of iterating with an estimator that assumes the same cosmology as the one derived from the data. This procedure could be quite time consuming, because one needs a large number of mock realizations for a given cosmology to optimize the estimator for this cosmology. We have found a way to do this efficiently by limiting the number of simulations to a few times the initial one. 

\subsection{Description of the method}
\label{subsec:description}
Our iterative procedure starts with a first calculation of the correlation function using the Landy-Szalay estimator. We then fit the resulting correlation function with a model that has considerable freedom on the general broadband shape, so that it is essentially sensitive to the location of the acoustic peak, as used in the BOSS analysis~[\cite{anderson}]:

\begin{equation}
\xi_{\mathrm{data}}(s) =  b^2~\xi_{\mathrm{theory}}(\alpha s) + a_0 + \frac{a_1}{s} + \frac{a_2}{s^2} \; ,
\label{eq:fit_model}
\end{equation}

\noindent where $\xi_\mathrm{theory}$ is the theoretical linear model from \cite{EisensteinHu}, $b$ the constant galaxy dark-matter bias factor, and $a_0$, $a_1$, and $a_2$ are nuisance parameters.

From this fit, we obtain the first iteration of the dilation scale parameter, $\alpha$, that characterizes the location of the peak: 

\begin{equation}
\alpha = \left(\frac{D_V}{r_s}\right) / \left(\frac{D_V}{r_s} \right)_f   \; ,
\label{eq:alpha_definition}
\end{equation}

\noindent  where $r_s$ is the comoving sound horizon at decoupling. The subscript $f$ means that the quantity is calculated using our fiducial cosmology, for which $r_s = 157.42$~Mpc; $D_V(z)$ is the spherically averaged distance to redshift $z$ and is defined by~[\cite{mehta}]:

\begin{equation}
D_V(z)=\left( (1+z)^2\frac{D_A^2(z)cz}{H(z)}\right)^{1/3} \; .
\end{equation}

The parameter $\alpha$ is unity if the actual cosmology matches the fiducial one. The result of this fit is a first estimate of the cosmological model suggested by the data, labelled by $\alpha_0$. This is actually the result of the standard analysis with the Landy-Szalay estimator.

We then perform a large number of realizations of mock catalogues with the same geometry as the data, for various values of $\alpha$ around $\alpha_0$. For this work we use lognormal simulations that can be quickly generated. 

For the set of simulations corresponding to a given input $\alpha_k$, one can find the coefficients $\vec{c}_k$ by minimizing the $\chi^2$ defined in Eq.~\ref{eq:chi2}. The resulting correlation function, $\hat{\vec{\xi}}_j(\vec{c}_k)$, for the realization $j$ is given by Eq.~\ref{eq:opt_est_definition}.  We compute the average  residual bias $\vec{B}_k$ of the  correlation function with respect to the theory $\vec{\xi}_\mathrm{theory}(\alpha_k)$:

\begin{equation}
\vec{B}_k = \left\langle \hat{\vec{\xi}}_j(\vec{c}_k) - \vec{\xi}_\mathrm{theory}(\alpha_k) \right\rangle_j \; .
\end{equation}
The covariance matrix $N_{\mathrm{opt}}(\alpha_k)$ is obtained from the fluctuations  of the same $n$ realizations:

\begin{equation}
N_\mathrm{opt}(\alpha_k) = 
\left\langle	\left[\vec{\xi}_j(\vec{c}_k) - \bar{\vec{\xi}}(\vec{c}_k)	\right] 
			\left[\vec{\xi}_j(\vec{c}_k) - \bar{\vec{\xi}}(\vec{c}_k)	\right]^T    \right\rangle_j  \; .
\end{equation}

Hereafter we redefine the process of applying the estimator corresponding to $\alpha_k$ to a data sample in two steps: 

\begin{itemize}
\item use Eq.~\ref{eq:opt_est_definition} with coefficients $ \vec{c}_k$ to calculate $\hat{\vec{\xi}}_\mathrm{data}(\vec{c}_k)$
\item add the residual bias of the estimator, $\vec{B}_k$.
\end{itemize}

We can now proceed with the iterative procedure. Since the first iteration value, $\alpha = \alpha_0$, is not exactly one of the nine available $\alpha_k$, we apply the estimator corresponding to the closest two values of $\alpha_0$,  $\alpha_\mathrm{lo}$, and $\alpha_\mathrm{hi}$, to the data, and we interpolate between the two resulting correlation functions:
\begin{equation}
\vec{\xi}^\mathrm{opt} (\alpha_0) = (1-t)~\vec{\xi}(\alpha_\mathrm{lo})+ t~\vec{\xi}(\alpha_\mathrm{hi}) \; ,
\label{eq:interpolation_of_xi}
\end{equation}
where $t = (\alpha_0-\alpha_\mathrm{lo}) / (\alpha_\mathrm{hi}-\alpha_\mathrm{lo})$.
Similarly, the covariance matrix can be written as a function of the two covariance matrices $N_{\mathrm{opt}}(\alpha_\mathrm{lo})$ and $ N_{\mathrm{opt}}(\alpha_\mathrm{hi})$ as

\begin{equation}
\begin{split}
N_{\mathrm{opt}}(\alpha)  = & \  (1-t)^2 N_{\mathrm{opt}}(\alpha_\mathrm{lo}) + t^2 N_{\mathrm{opt}}(\alpha_\mathrm{hi}) \\   & + t(1-t)~C_\mathrm{opt}(\alpha_\mathrm{lo},\alpha_\mathrm{hi}) \; ,
\end{split}
\label{eq:interpolation_of_cov}
\end{equation}
where $C_\mathrm{opt}(\alpha_\mathrm{lo},\alpha_\mathrm{hi})$ is the cross-covariance between $\vec{\xi}^\mathrm{opt} (\alpha_\mathrm{lo})$ and $\vec{\xi}^\mathrm{opt} (\alpha_\mathrm{hi})$ given by

\begin{multline}
C_\mathrm{opt}(\alpha_\mathrm{lo},\alpha_\mathrm{hi}) = \\
\Big\langle \left[ \vec{\xi}^\mathrm{opt}(\alpha_\mathrm{lo}) - 				
	 				\bar{\vec{\xi}}^\mathrm{opt}(\alpha_\mathrm{lo}) 		\right]			
	 		\left[ \vec{\xi}^\mathrm{opt}(\alpha_\mathrm{hi}) - 
	 				\bar{\vec{\xi}}^\mathrm{opt}(\alpha_\mathrm{hi})		\right]^T  \\ 
		+ 	\left[ \vec{\xi}^\mathrm{opt}(\alpha_\mathrm{hi}) - 						
				 	\bar{\vec{\xi}}^\mathrm{opt}(\alpha_\mathrm{hi})		\right]
			\left[ \vec{\xi}^\mathrm{opt}(\alpha_\mathrm{lo}) -
					\bar{\vec{\xi}}^\mathrm{opt}(\alpha_\mathrm{lo})		\right]^T \Big\rangle_s \; .
\end{multline}

Finally, we fit the correlation function (Eq.~\ref{eq:interpolation_of_xi}) with the template (Eq.~\ref{eq:fit_model}) using the covariance matrix (Eq.~\ref{eq:interpolation_of_cov}), which yields a new value $\alpha_1$ at the second iteration. We then iterate until the estimated $\alpha_i$ varies less than a given quantity between two successive iterations. In practice, convergence is achieved after a few iterations. 

\subsection{Implementation of the method}

We have used lognormal simulations to calculate the estimator for each cosmology (through different values of $\alpha$).
There are several reasons for this choice: the first one is that the theoretical input correlation function is known for such simulations, allowing $B_k$ to be calculated in a straightforward manner. The second reason is that such mock catalogues are easy and fast to produce, allowing for a large number of realizations. The method could, however, be adapted to any other kind of simulations.
 
 We perform a large number of realizations of mock catalogues with the same DR9 geometry as the data, for various values of $\alpha$ around $\alpha_0$. 
For this work we simulated 120 realizations of 9 different cosmologies such that the dilation parameter $\alpha$ covers the range $[0.96,1.04]$ in steps of width $0.01$. For each realization we used 271,000 galaxies and a random catalogue that is 15 times larger. We then applied the iterative optimal estimator described in the previous section until the estimated $\alpha_i$ vary less than $\Delta \alpha=0.0001$ (achieved after just a few iterations).

 The method proposed requires  modest extra CPU time compared to the standard analysis. In general, the pair counting is one of the most time-consuming processes of any galaxy correlation analysis, and it needs to be done for any kind of estimator. The extra CPU time comes from pair counting of the mock catalogues with different cosmologies. The iterative procedure required to have an unbiased estimator needs at least two extra sets of mock catalogues with different BAO peak positions. In the case considered here (9 cosmologies) the production and pair counting of the lognormal mock catalogues required a few days on a single desktop machine. As shown in the following sections, the improvement in terms of error on $\alpha$ is around 20\% showing that it is worth the effort.

\subsection{Performance on simulations}
\label{subsec:performances}

In this section, we investigate the properties of the iterative optimal estimator on mock catalogues. We start with the lognormal simulations used to optimize the estimator and show that we derive an estimator that is indeed independent of the input cosmological model. We obtain a 20-25\% increase in the accuracy of the $\alpha$ parameter with these simulations. We also test our estimator on simulations that are different from the lognormal ones used to optimize it. These are more realistic simulations than the lognomal simulations; they were produced in the framework of the SDSS-III/BOSS galaxy clustering working groups.

\subsubsection{Lognormal mock data}

\begin{figure}
   \centering
  \resizebox{\hsize}{!}{\includegraphics{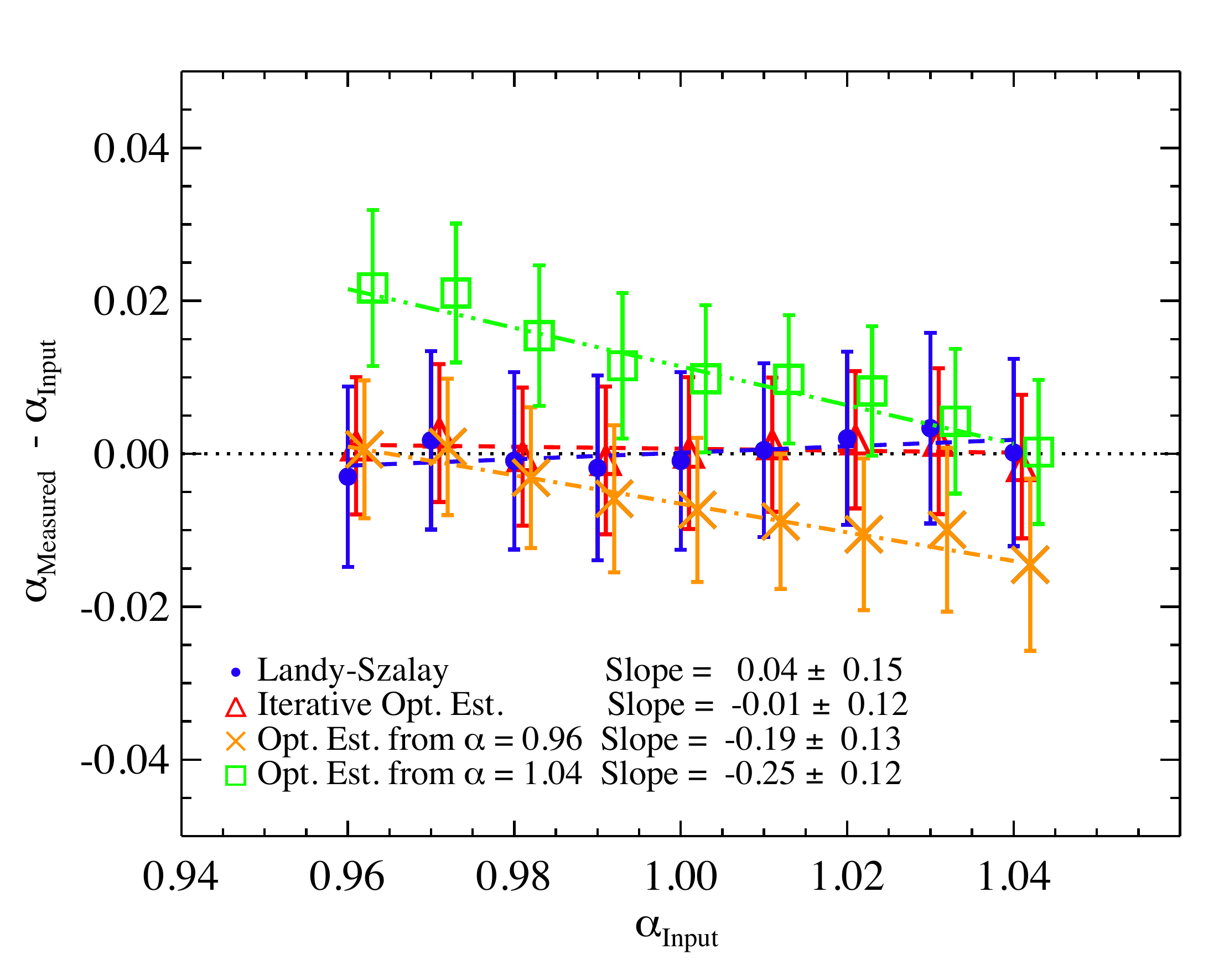}}
\caption{ Bias on $\alpha_\mathrm{Measured}$ as a function of $\alpha_\mathrm{Input}$ (from lognormal simulations) for Landy-Szalay (blue points), Iterative optimal (red triangles) and two non-iterative minimum-variance estimators with $\alpha=0.96$ (orange crosses) and $\alpha=1.04$ (green squares). The error bars indicate the RMS among the lognormal simulations. Simple linear functions were fitted to the points in each case with corresponding slopes shown in the legend. A strong bias can be seen for the non-iterative estimators while the iterative optimal and Landy-Szalay estimators are not biased. (Coloured version of the figure available online).
}
       \label{fig:res_alphabias}%
\end{figure}

As an illustration, we first considered what happens when we do not use the iterative procedure. The nine different sets of 120 lognormal simulations provide nine different optimal estimators, defined by $\vec{c}_k$ and $\vec{B}_k$. 
We choose two of them to apply to the nine sets of simulations \textit{without} the iterative procedure. We fit the resulting correlation function with the template of Eq. \ref{eq:fit_model} to obtain the scale parameter, $\alpha_\mathrm{Measured}$. 
In Fig.~\ref{fig:res_alphabias}, $\alpha_\mathrm{Measured}$ is averaged over the 120 simulations with the same given $\alpha_\mathrm{Input}$, and its difference with $\alpha_\mathrm{Input}$ is plotted versus $\alpha_\mathrm{Input}$.
The non-iterative estimators do not recover $\alpha_\mathrm{Input}$. This result occurs because the peak-shaped (Fig.~\ref{fig:estimator_bias}) residual bias correction, $B_k(\alpha,r)$ slightly shifts the peak position of the data to the left ($\alpha=0.96$) or to the right ($\alpha=1.04$). As emphasized by the linear fits in the figure, the residual bias increases with the difference $\alpha - \alpha_\mathrm{Input}$. 

Figure ~\ref{fig:res_alphabias} also demonstrates what happens with the iterative optimal estimator. The iterative procedure indeed removes the residual bias, since the iterative optimal estimator appears to be unbiased. 
The error bars in the figure are RMS of the values of $\alpha_\mathrm{Measured}$ for the 120 different simulations. The optimal estimator gives smaller RMS than the Landy-Szalay estimator. This result is confirmed in Fig.~\ref{fig:resolution}, which shows this RMS as a function of $\alpha_\mathrm{Input}$. The gain obtained with the optimal estimator is $\approx$22\% relative to the Landy-Szalay estimator (this number is not a general one; the precise value depends on the geometry of the survey)
 leading to similar improvement on subsequent cosmological constraints.

\begin{figure}
   \centering
   \resizebox{\hsize}{!}{\includegraphics{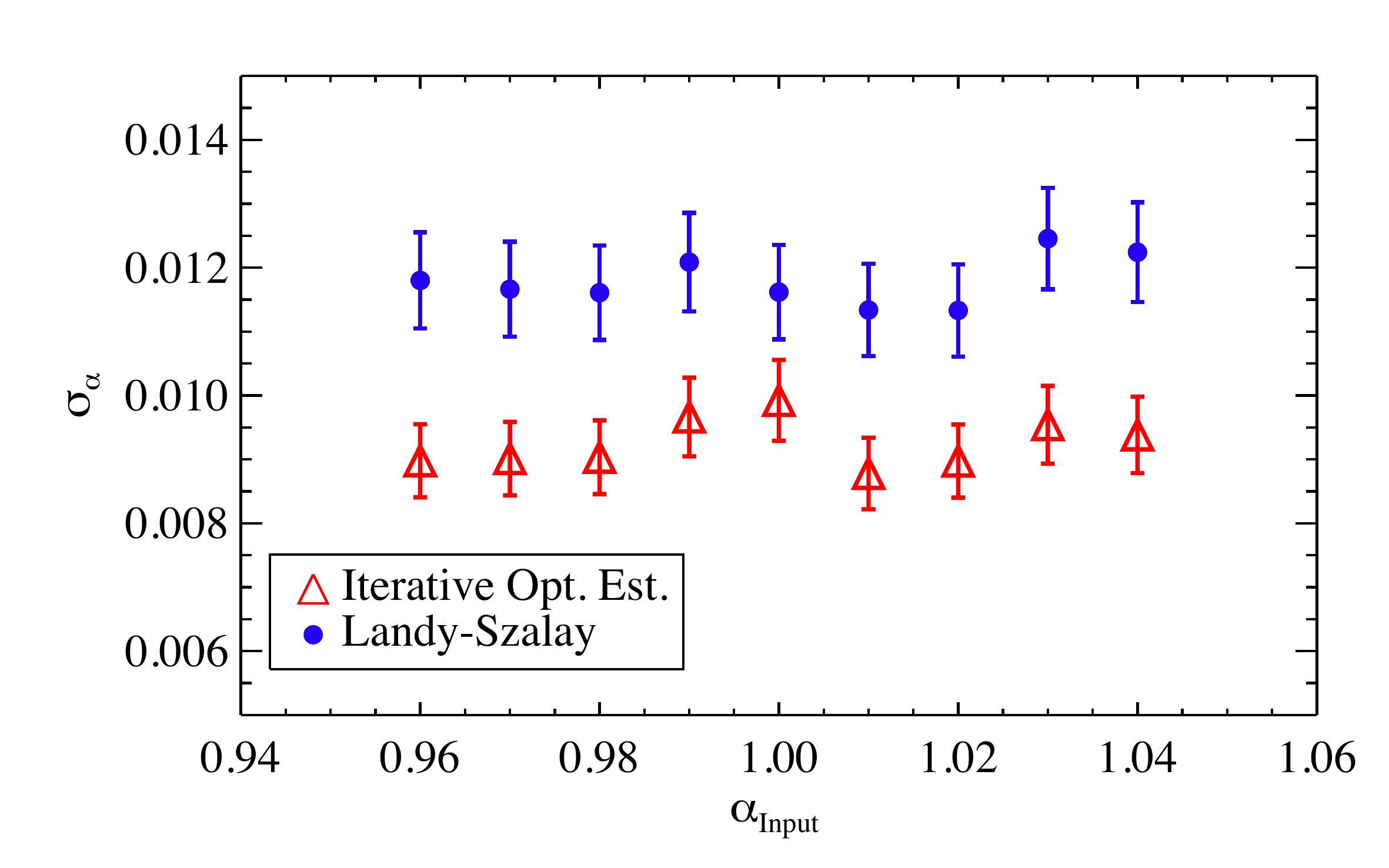}}
   \caption{ Error on $\alpha_\mathrm{Measured}$ for the iterative optimal (red triangles) and Landy-Szalay (blue points) estimators for the nine sets of realizations with different $\alpha_\mathrm{Input}$. Error bars indicate the RMS among the lognormal simulations. The gain using the iterative optimal estimator is obvious. (Coloured version of the figure available online).}
              \label{fig:resolution}%
\end{figure}

\subsubsection{PTHalos and LasDamas mock data}

The studies in the previous section were performed by applying the iterative optimal estimator to the lognormal simulations that were used to optimize the estimator. We repeated the calculations using two other sets of mock catalogues that have very similar geometries, based on the BOSS DR9 footprint [\cite{anderson}]. 

The first set is based on the second-order Lagrangian perturbation theory matter field [\cite{scoccimaro}] and halo occupation function named PTHalos [\cite{manera}].  A total of 610 realizations were produced with $h=0.7$, $\Omega_m=0.274$, $\Omega_\Lambda=0.726$, $\Omega_bh^2=0.0224$, $\sigma_8=0.8$, and $n_s=0.97$\footnote{For this work we used 598 of the 610 realizations available. }. 
Since the fiducial cosmology used to compute the comoving distances of the galaxies is slightly different than the one used in mock catalogues, $\alpha$  (Eq.~\ref{eq:alpha_definition}) is not expected to be 1 but 1.002.

The second set is even more realistic; it uses $N$-body simulations, named large suite of dark matter simulations (LasDamas) [McBride et al., in preparation]; developed within the SDSS I-II galaxy clustering working group for the DR7 LRG analysis. A total of 153 realizations were produced assuming a flat $\Lambda$CDM cosmology with $\Omega_b = 0.04$, $\Omega_m = 0.25$, $h = 0.7$, $n_s = 1.0$ and $\sigma_8 = 0.8$, for which $\alpha$ is expected to be 0.988.

\begin{figure}
   \centering
   {
   \includegraphics[width=0.45\textwidth]{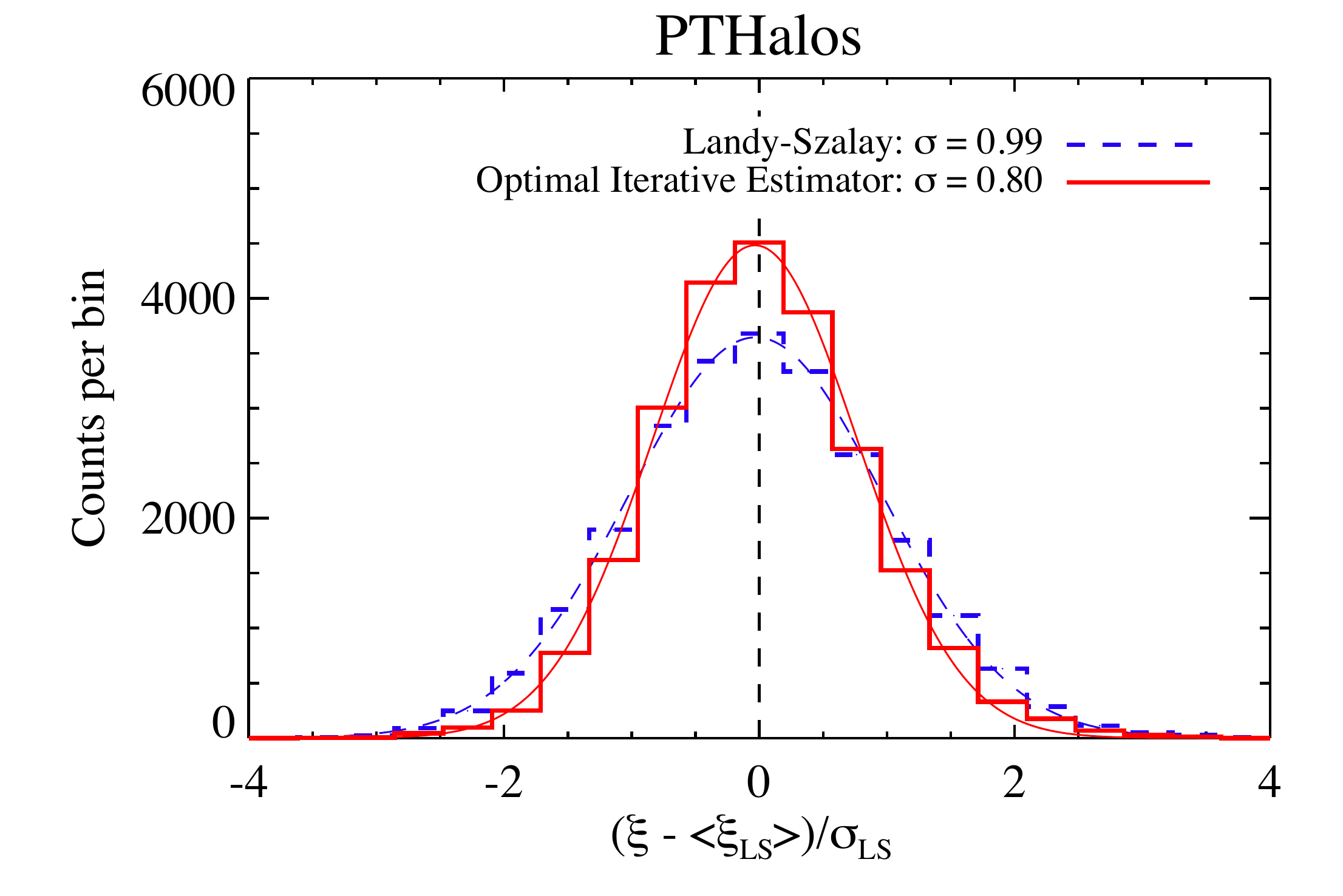}
	\includegraphics[width=0.45\textwidth]{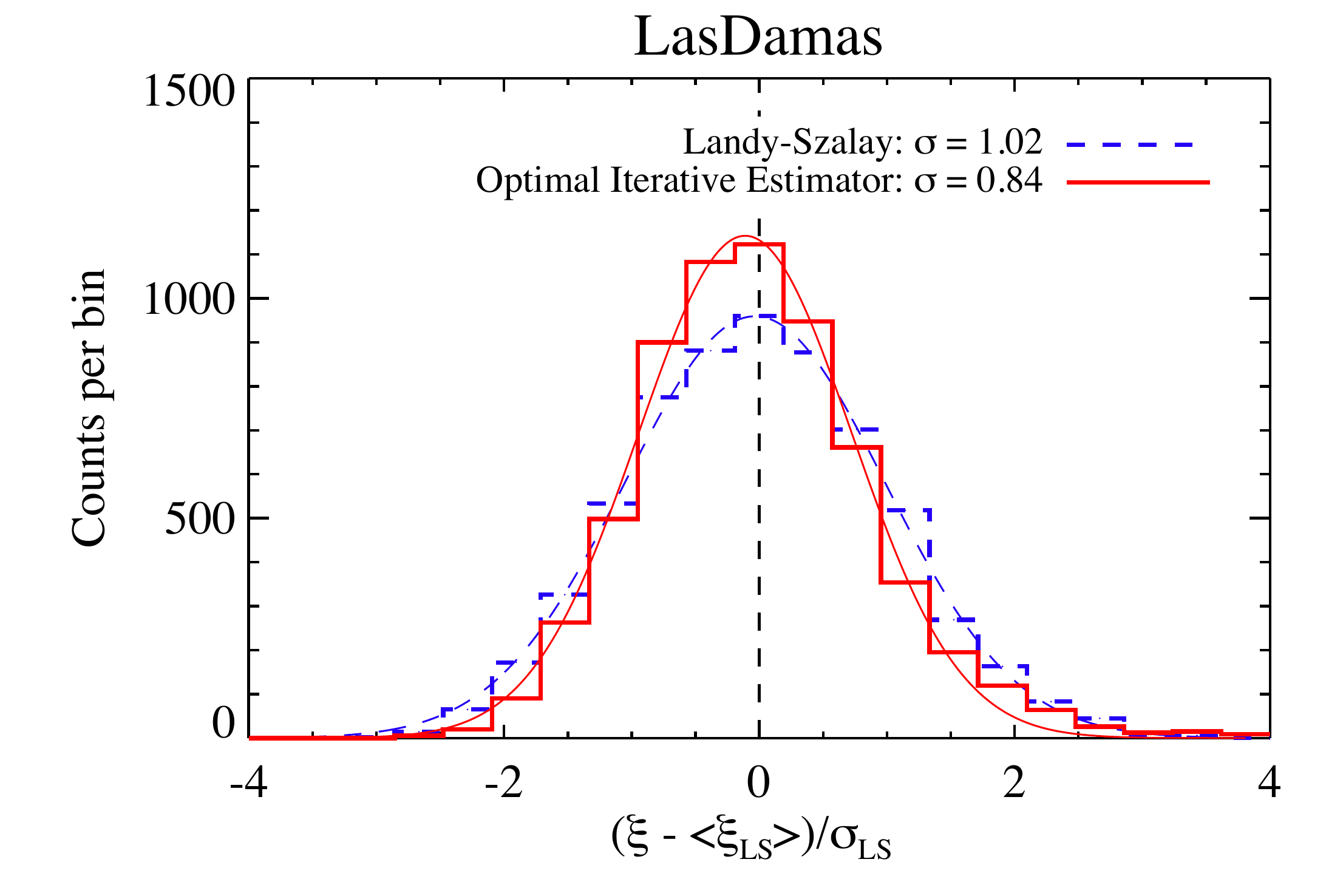}}
   \caption{``Pull" distribution of correlation functions measured with PTHalos (top) and LasDamas (bottom) mocks in the range $40<s<200~h^{-1}$ Mpc with the Landy-Szalay {\bf (dashed-blue) and the iterative optimal estimator (solid red}). The standard deviation of the Gaussian fit shows a smaller scatter for the latter estimator. 
   (Coloured version of the figure available online)
   }
              \label{fig:pull}%
\end{figure}

Figure ~\ref{fig:pull} shows the ``pull'' histogram of the correlation functions, i.e., the residuals of the correlation functions relative to the average Landy-Szalay correlation function, normalized to the empirical RMS of the Landy-Szalay estimator, $(\xi-\langle \xi_{LS}\rangle)/\sigma_{LS}$. By construction, the width of the pull distribution for the Landy-Szalay estimator is close to one for both sets of mock data, while it is 0.80 for the iterative optimal estimator (PTHalos) and 0.84 (LasDamas), which is similar to the 22\% gain on the error bars obtained with the lognormal simulations. 

This result is confirmed by Fig.~\ref{fig:cov_pthalos}, which shows the covariance and correlation matrices obtained with both estimators on the PTHalos mocks. The gain in the covariance matrix elements is obvious and not partially cancelled by an increase in the off-diagonal terms in the correlation matrix. Figure ~\ref{fig:cov_lasdamas} shows the same information for the LasDamas simulations. The matrices are noisier since we have fewer realizations, but the improvement is visible, and again the correlation does not change. A small increase in the covariance matrix is present between 40 and 60 $h^{-1}$ Mpc, which can also be seen in Fig.~\ref{fig:estimator_bias}. However, these scales are much smaller than the scales of interest here (BAO peak) where we indeed see a clear reduction of the covariance.

\begin{figure}
   \centering
    \includegraphics[width=0.49\columnwidth]{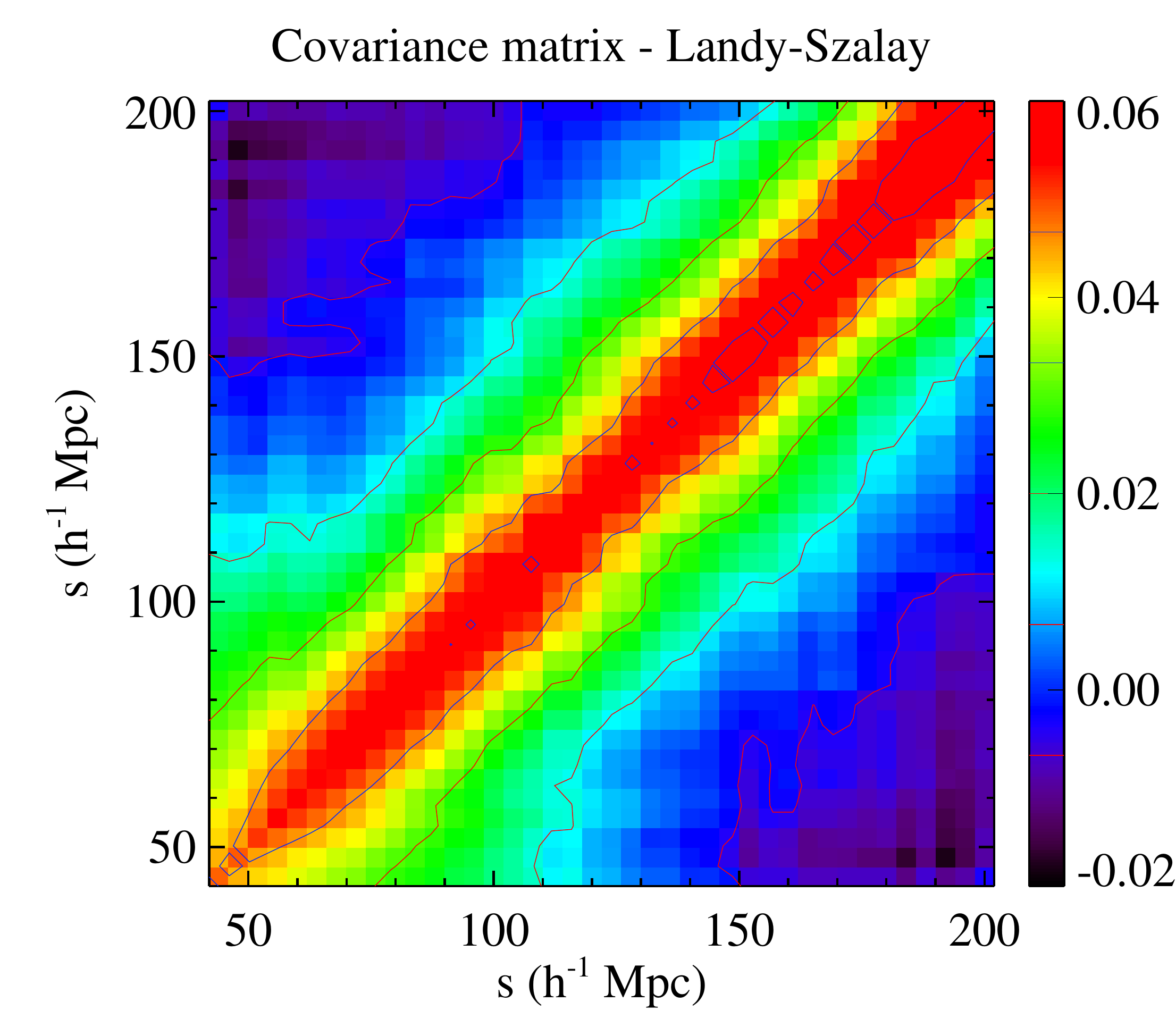}
   \includegraphics[width=0.49\columnwidth]{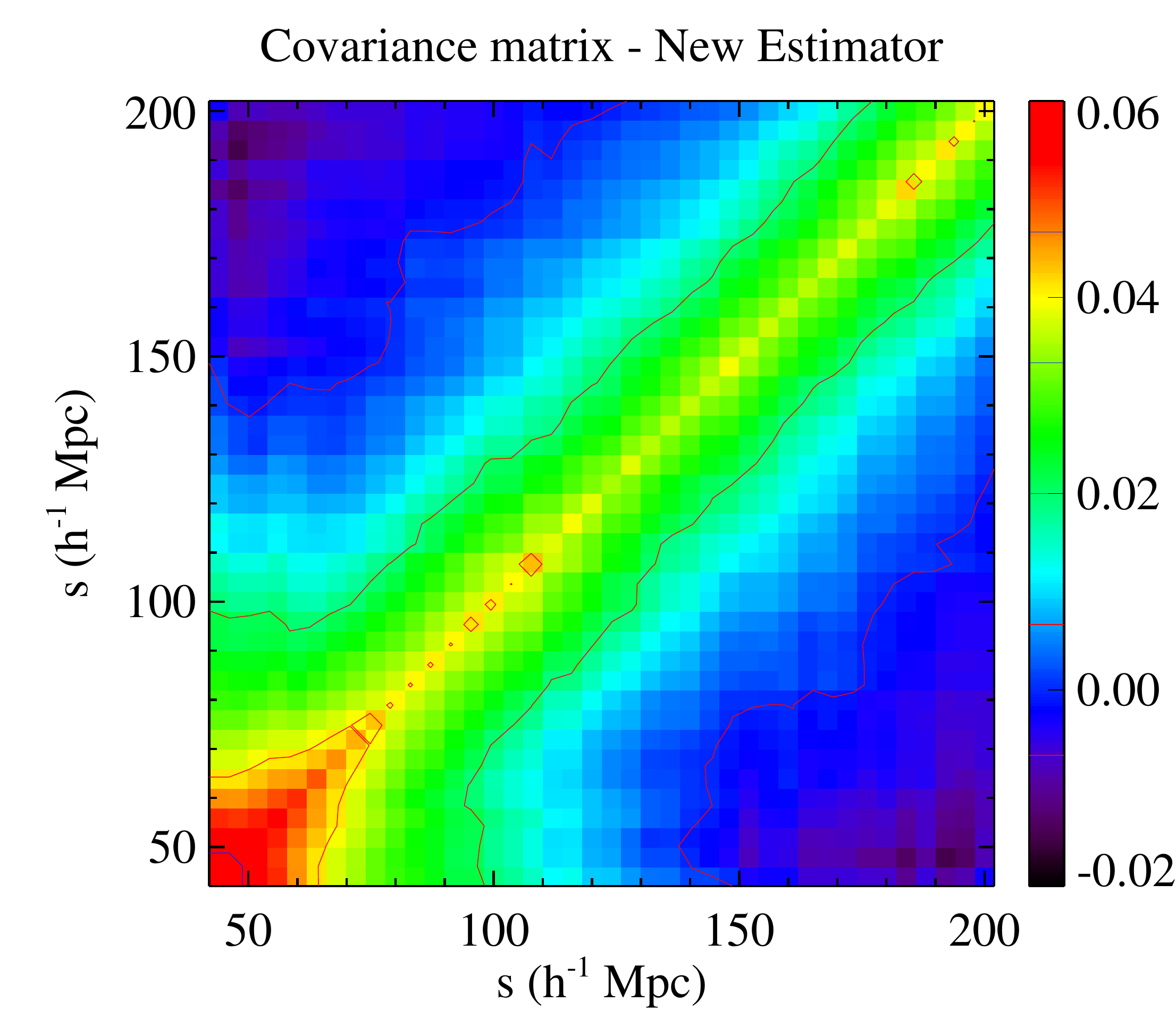}
   \includegraphics[width=0.49\columnwidth]{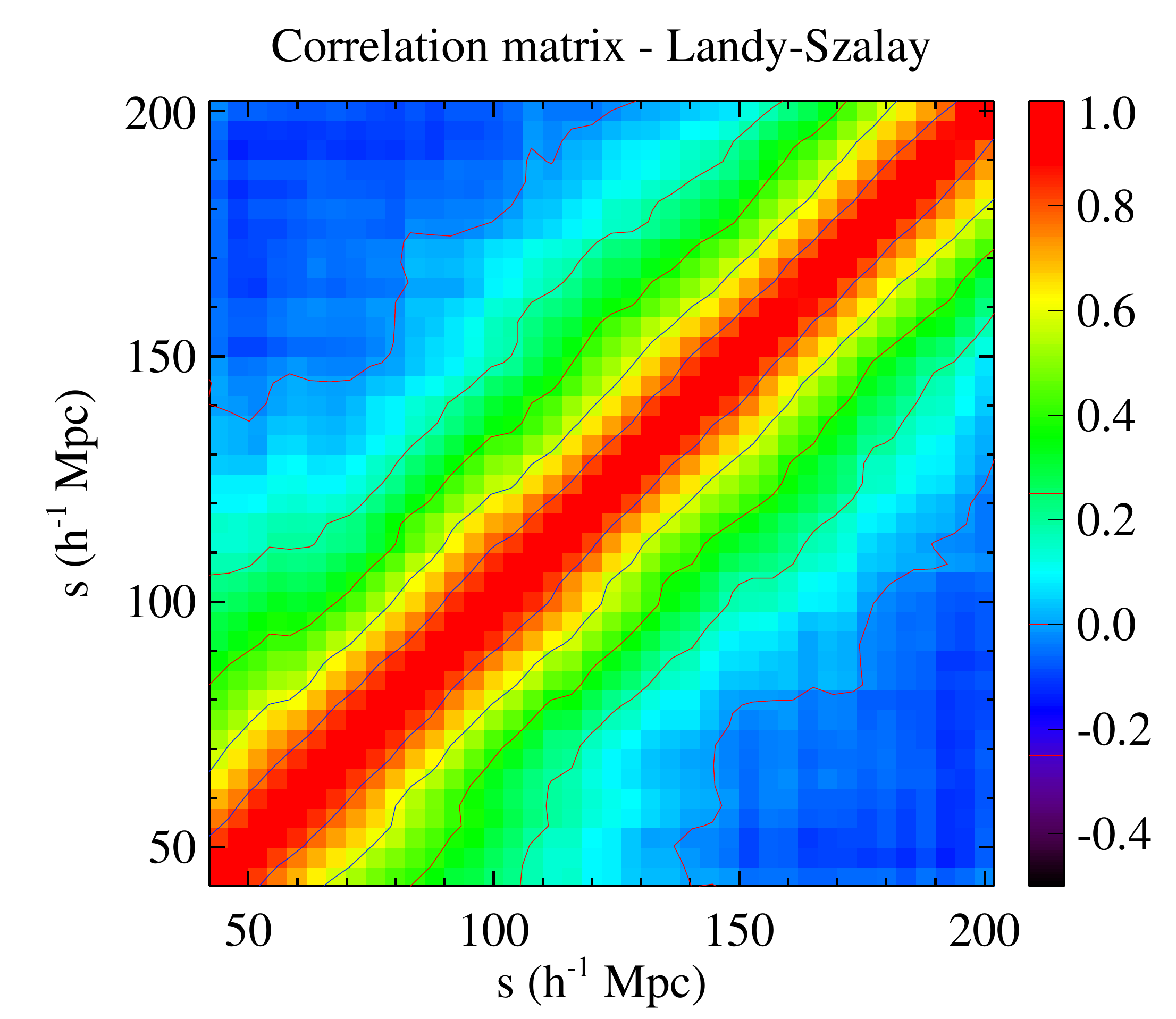}
   \includegraphics[width=0.49\columnwidth]{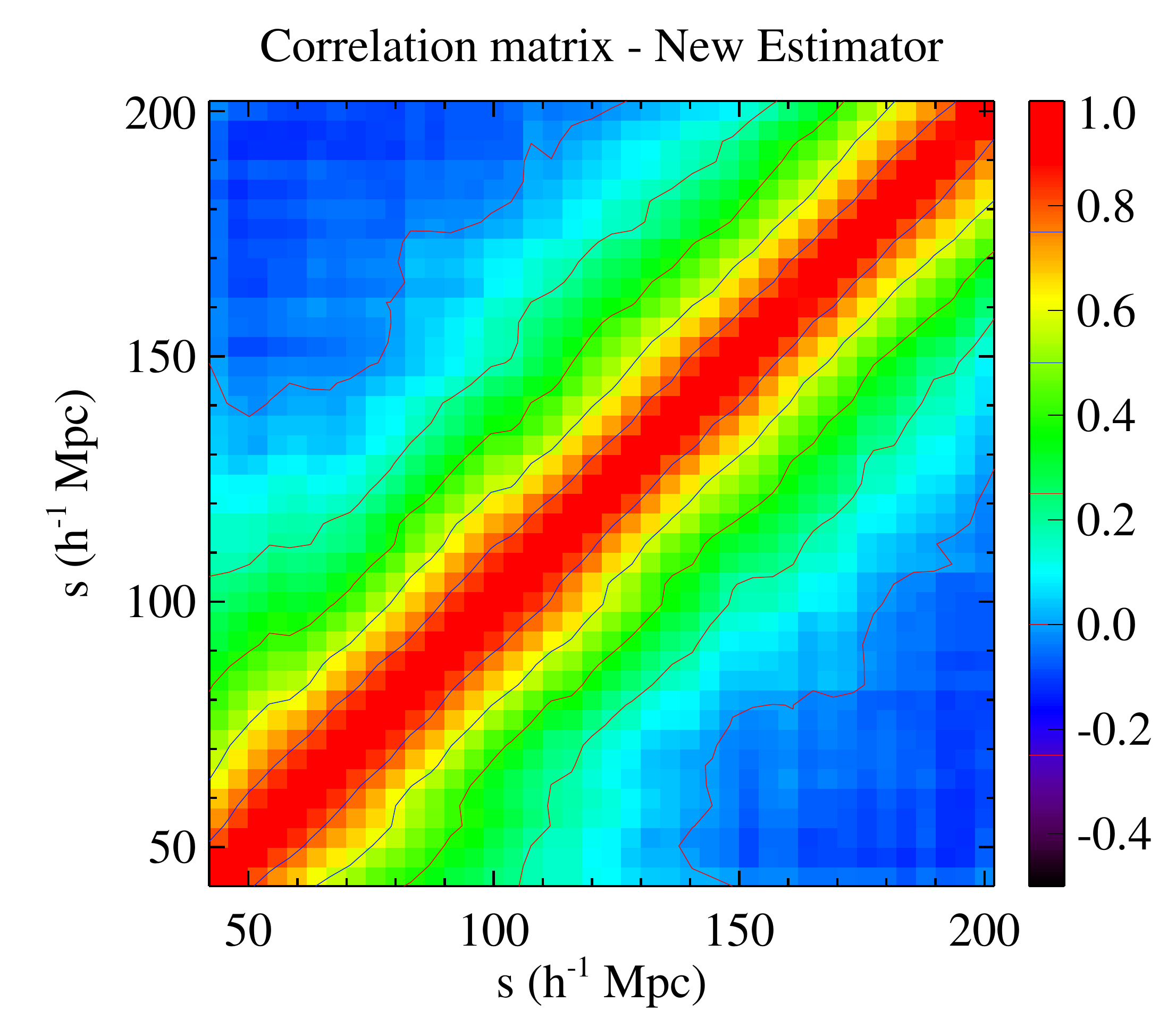}
    \caption{Covariance matrices times the square of the comoving distance (top panels) and correlation matrices (bottom panels) of PTHalos mock catalogues using Landy-Szalay (left panels) and the iterative optimal estimator (right panels). 
    (Coloured version of the figure available online)
    }
             \label{fig:cov_pthalos}%
\end{figure}
\begin{figure}
   \centering
    \includegraphics[width=0.49\columnwidth]{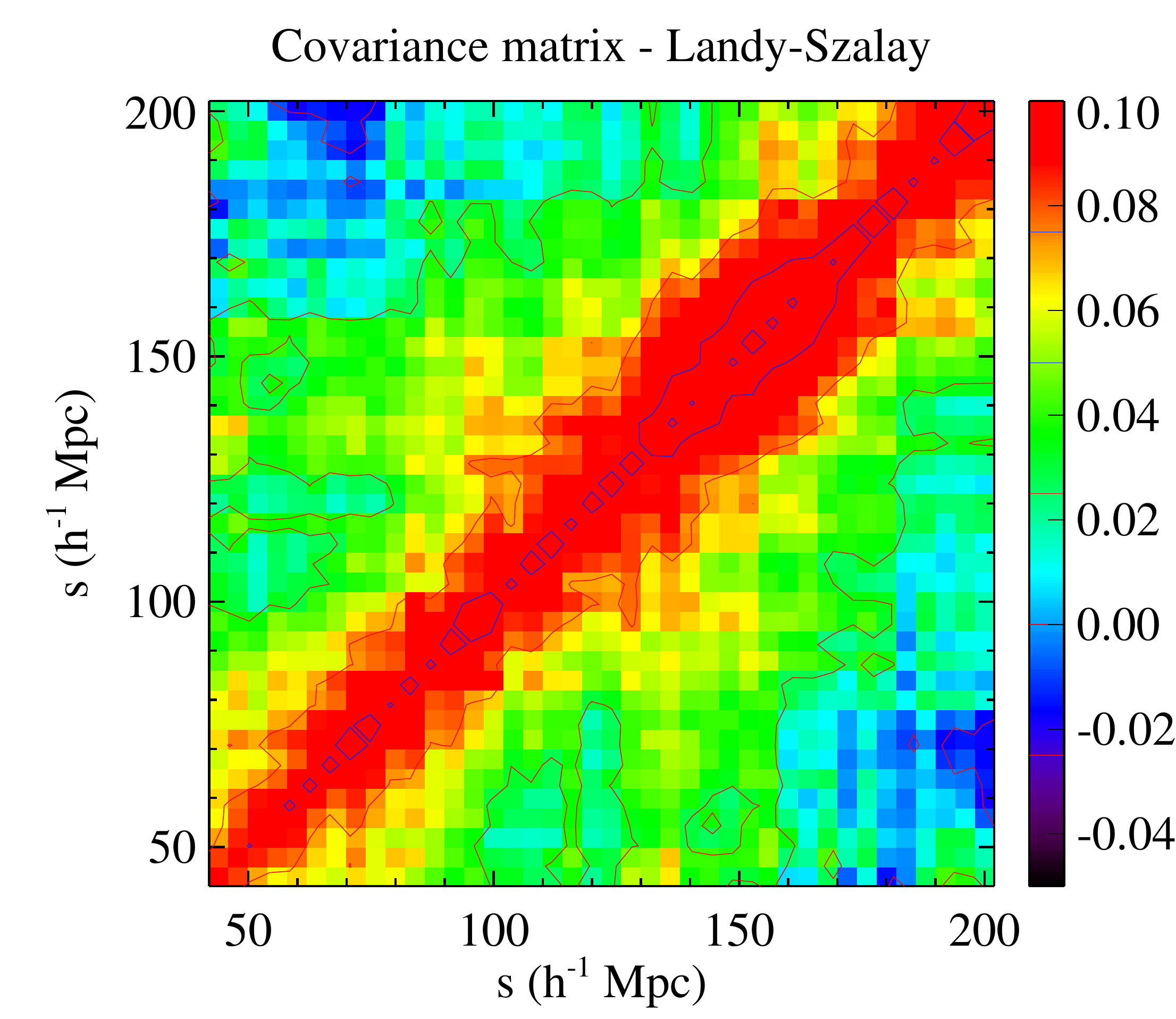}
   \includegraphics[width=0.49\columnwidth]{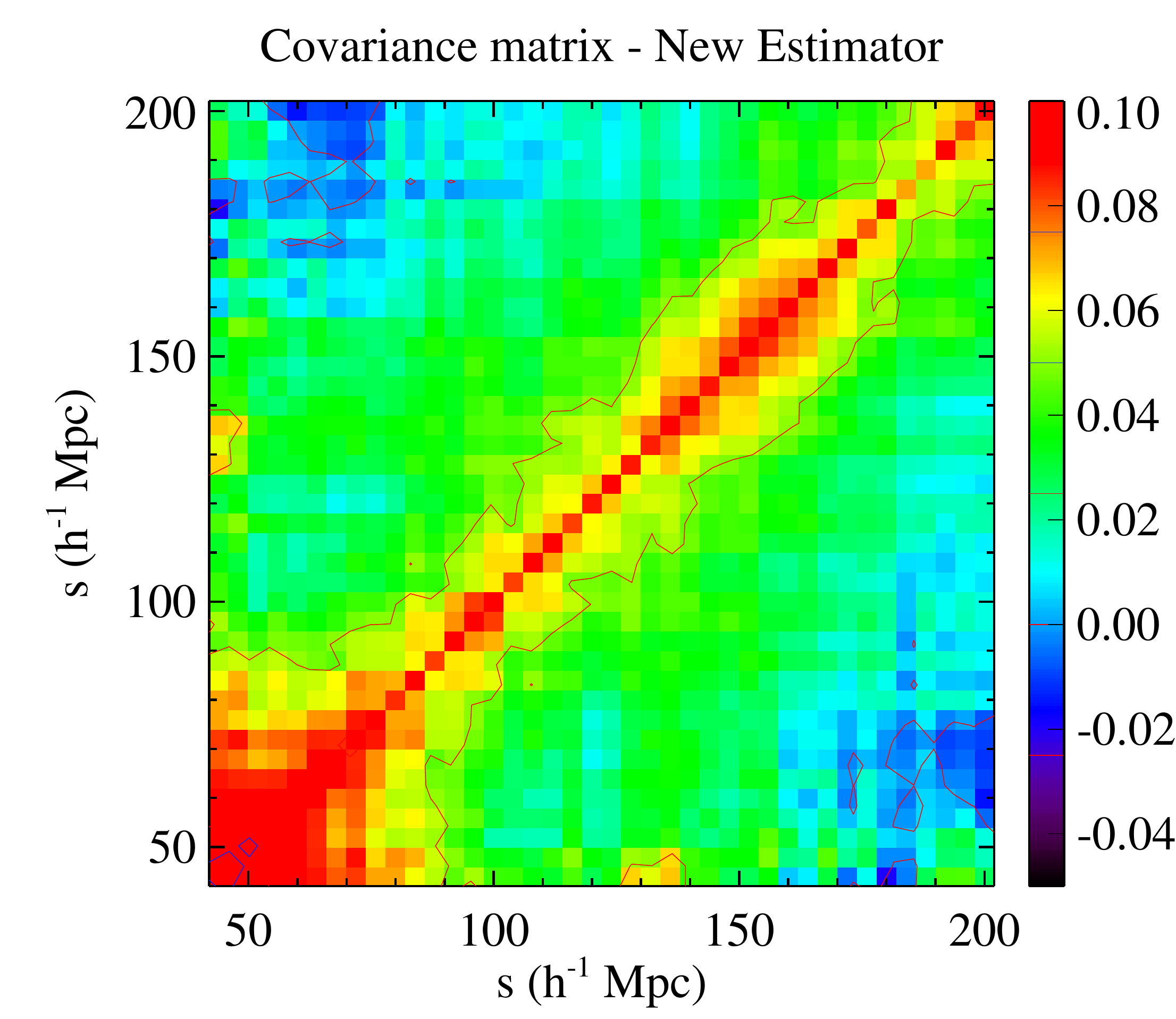}
   \includegraphics[width=0.49\columnwidth]{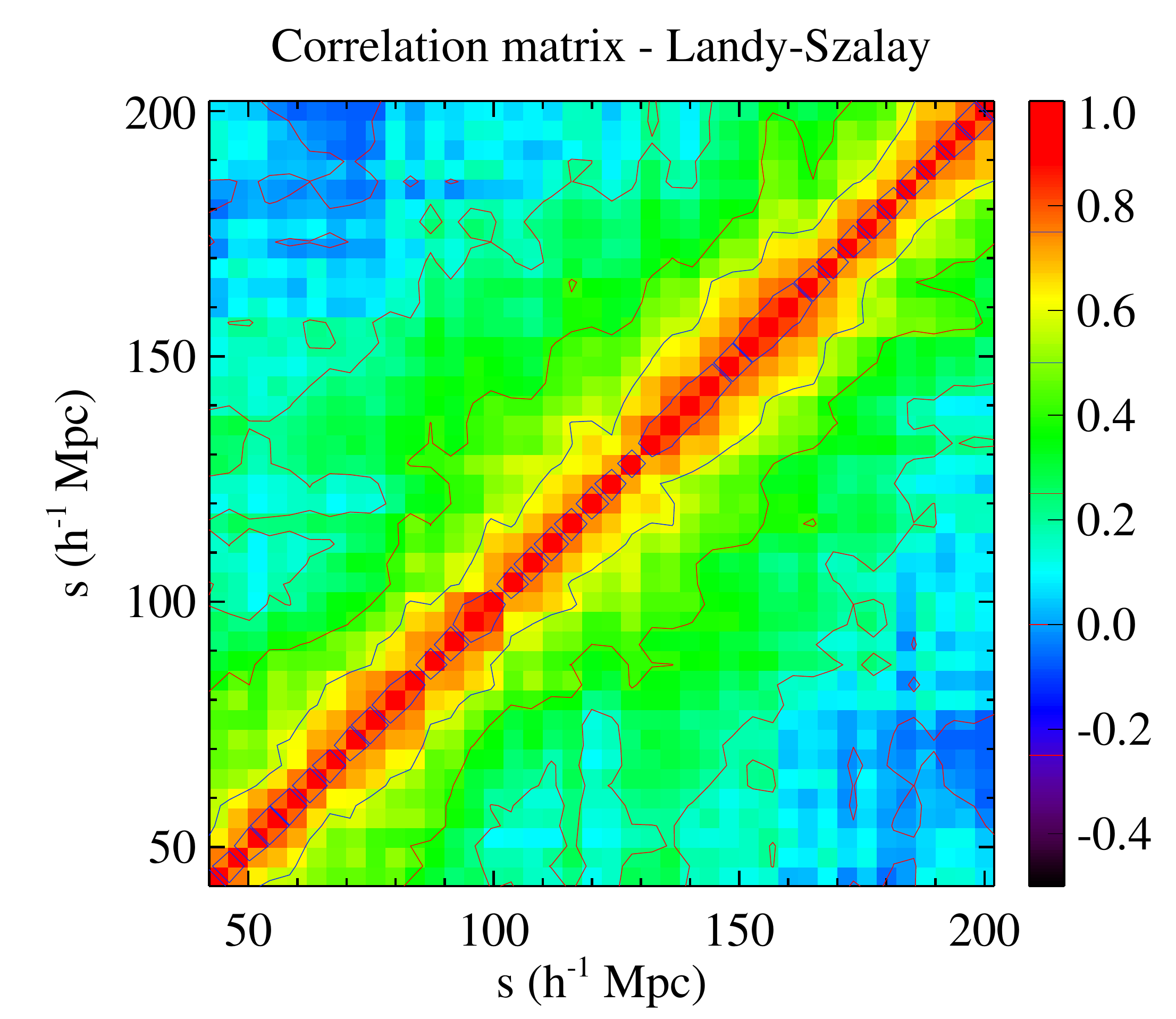}
   \includegraphics[width=0.49\columnwidth]{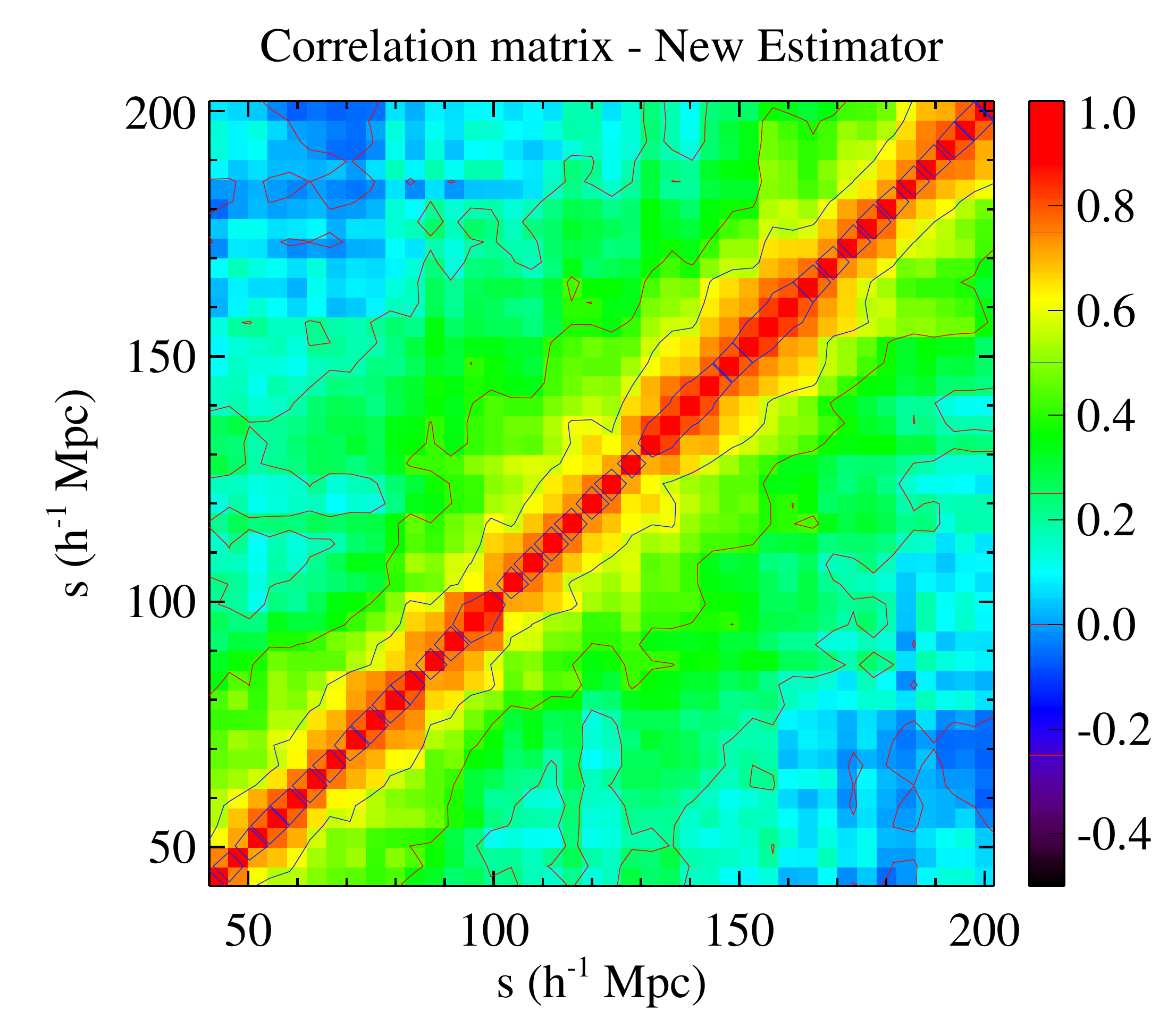}
	\caption{Same as figure \ref{fig:cov_pthalos} for LasDamas mock catalogues. 
	(Coloured version of the figure available online)
	}  
              \label{fig:cov_lasdamas}%
\end{figure}
Finally, Fig.~\ref{fig:hist_alpha} displays the improvement on the estimation of $\alpha$ obtained using the iterative optimal estimator. The scatter of $\alpha_\mathrm{Measured}$ with the optimal estimator is reduced relative to Landy-Szalay by 21\% for PTHalos and 17\% for LasDamas mocks. These gains are consistent with the observed ``pull" distribution (Fig.~\ref{fig:pull}) and confirm the gain observed with the lognormal simulations.

\section{Application to real data\label{data}}
\subsection{Data description}

 We apply our final estimator on two galaxy samples: the SDSS I-II  DR7 luminous red galaxy sample (LRG) [~\cite{eisenstein2001}] and the SDSS-III/BOSS DR9 CMASS [Padmanabhan, N. et al., in preparation].
 
Both surveys, SDSS-I-II and SDSS-III/BOSS use the same wide field and a dedicated telescope, the 2.5 m-aperture Sloan Foundation Telescope at Apache Point Observatory in New Mexico [\cite{Gunn2006}]. Those surveys imaged the sky at high latitude in the ugriz bands [\cite{Fukugita1996}], using a mosaic CCD camera with a field of view spanning 2.5 deg  [\cite{Gunn1998}].
The SDSS-I-II imaging survey is described  in \cite{abazajian2009}.  
This galaxy catalogue had been built based on the prescription in \cite{eisenstein2001}, selecting the most luminous galaxies since they are more massive and then more biased with respect to the dark-matter density field. More details about the construction of the catalogue can be found in \cite{kazin2010}. The galaxies in the LRG sample have redshifts in the range $0.16 < z < 0.47$ and a density of  about $10^{-4}h^3$ Mpc$^{-3}$.

The BOSS imaging survey data are described in \cite{Aihara2011}, the spectrograph design and performance in \cite{Smee}, and the spectral data reductions in Schlegel, D. et al. 2012, in preparation and \cite{bolton}.  A summary of BOSS can be found in \cite{dawson}.
The SDSS Data Release 9 [\cite{ahn2012}] CMASS sample of galaxies is constructed using an 
extension of the selection algorithm of DR7 LRG sample in order to detect fainter and bluer massive galaxies lying in the redshift range $0.43<z<0.7$. The final density of this sample is $3 \times 10^{-4}h^3$ Mpc$^{-3}$. A more detailed explanation of the target selection is given in Padmanabhan, N. et al., in preparation.
\subsection{DR7}

We compared the iterative optimal estimator to the Landy-Szalay estimator for estimating the spherically averaged two-point correlation function of the SDSS DR7 [\cite{abazajian2009}] LRG sample (Fig.~\ref{fig:dr7_fit}). 
To estimate the correlation function, we used a random catalogue that is 15 times larger than the data sample. The coefficients $\vec{c}_k$ and residual biases $B_k(\vec{r})$ for the iterative estimator were obtained using the nine sets of lognormal simulations as described in section \ref{subsec:description}. The covariance matrix of the data correlation function for both estimators comes from the 153 realizations of the LasDamas mocks. The top panels of Fig.~\ref{fig:dr7_fit} show the resulting correlation function for both estimators. The error bars obtained for the iterative estimator (right panel) are smaller than for the Landy-Szalay  estimator (left panel), but both curves are consistent with each other.   The error bars in the figure are the diagonal terms of the covariance matrix. These plots do not show the correlation between different separation bins (see Fig.\ref{fig:cov_pthalos}). As a consequence, certain points show small offsets from the fit that are not significant when the full covariance matrix is considered. The $\chi^2$/d.o.f = 28.4/35 and $p=0.78$  for Landy-Szalay and $ \chi^2$/d.o.f = 29.9/35 and $p=0.71$ for the optimal estimator, indicating a good fit in both cases.

The correlation functions were fitted as for the mock catalogues, using the template defined by Eq.~\ref{eq:fit_model}. The resulting values of $\alpha_\mathrm{Measured}$ are compatible with unity, and the error for the optimal estimator is lower by 31\%, as shown in Table~\ref{tab:alphas}. This improvement is larger than the 17\% improvement on the mean error observed on mock data, but it is consistent with the scatter of the errors (Fig.~\ref{fig:scatter_error}, left). 

In Fig.~\ref{fig:scatter_alpha} (left) we use both estimators to compare $\alpha_\mathrm{Measured}$ for the DR7 LRG sample and LasDamas realizations. The DR7 measurement is well inside the LasDamas cloud, since very close to the mean.

Another way to improve the measurement accuracy of the BAO peak is through the reconstruction technique, where galaxies are slightly displaced so that the density field is as it should be without non-linear structure growth effects [\cite{eisenstein2007}]. \cite{xu2012} used the reconstruction technique on the  DR7 LRG sample. Before reconstruction they obtained $\alpha = 1.015\pm0.044$, and after reconstruction $\alpha=1.012\pm0.024$, an improvement of errors of 45\%. Our estimator, with a 31\% improvement, yields $\alpha=1.008\pm0.018$, which is consistent with the reconstruction result. This comparison shows that it is possible to gain in accuracy in two independent ways, and the combination of both methods is expected to provide  even better constraints on cosmological parameters.

\begin{figure*}[t]
   \centering
  		{\includegraphics[width=0.45\textwidth]{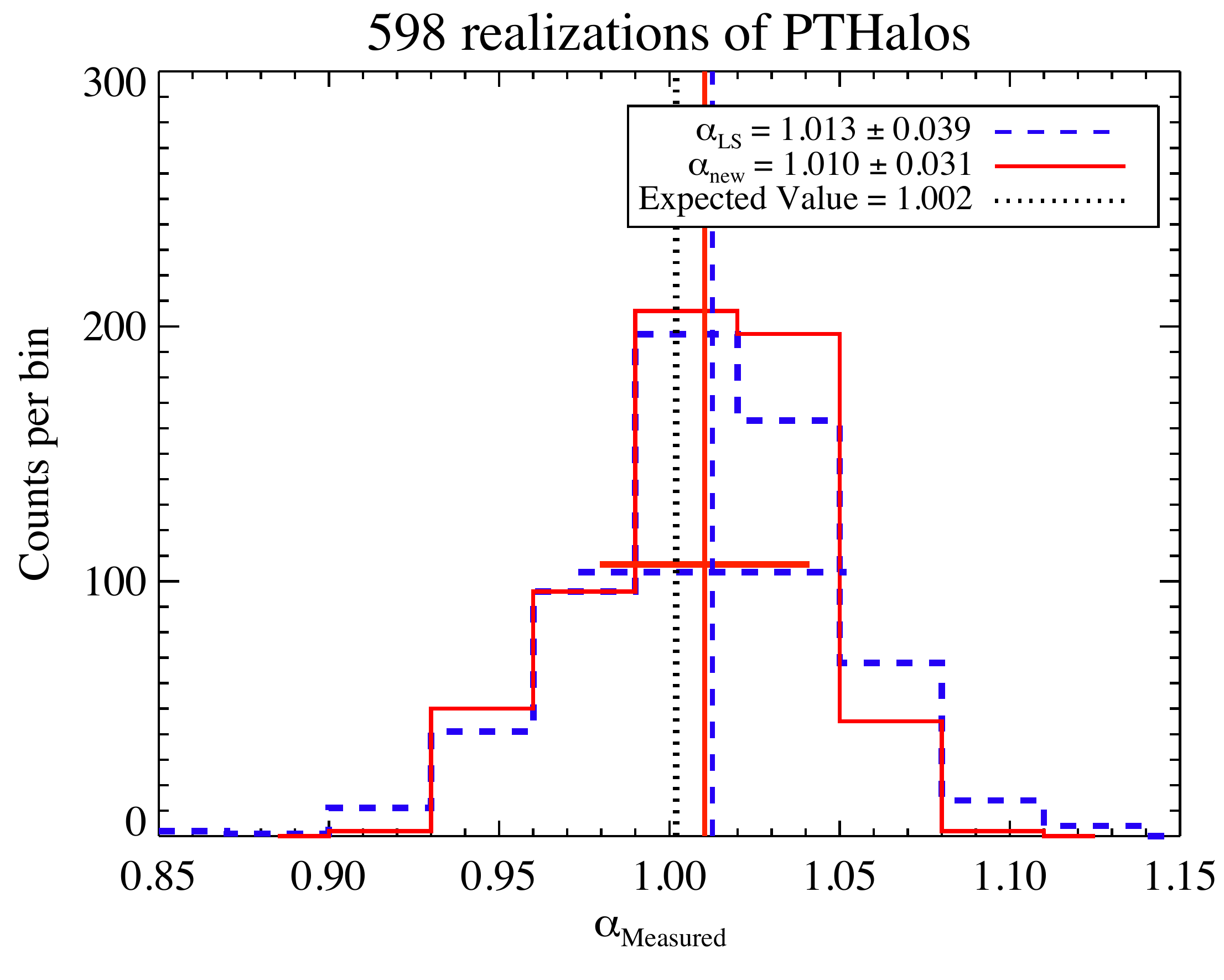}
    \includegraphics[width=0.45\textwidth]{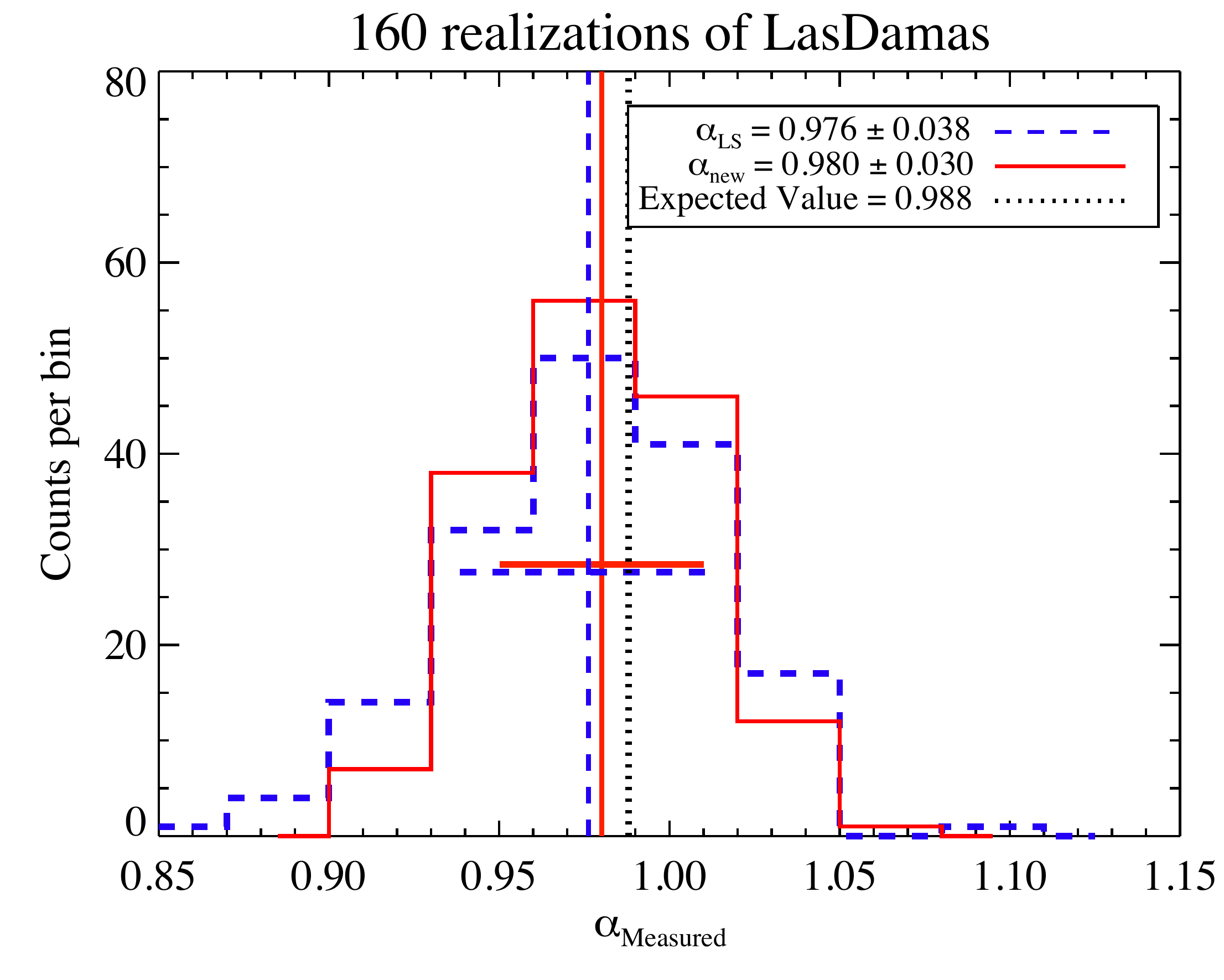}}
   \caption{ Histogram of $\alpha_\mathrm{Measured}$ for the PTHalos  (left) and  LasDamas (right) realizations using the Landy-Szalay (dashed blue) and the iterative optimal estimators (solid red). The average values over the realizations, shown in the legend, are represented as vertical lines with corresponding colours and linestyles, while the RMS of the histograms are shown as horizontal thick lines (corresponding colours). The expected theoretical value is shown as a vertical black dotted line.
   (Coloured version of the figure available online).
   }
              \label{fig:hist_alpha}%
\end{figure*}
\begin{figure*}
   \centering
  {\includegraphics[width=0.49\textwidth]{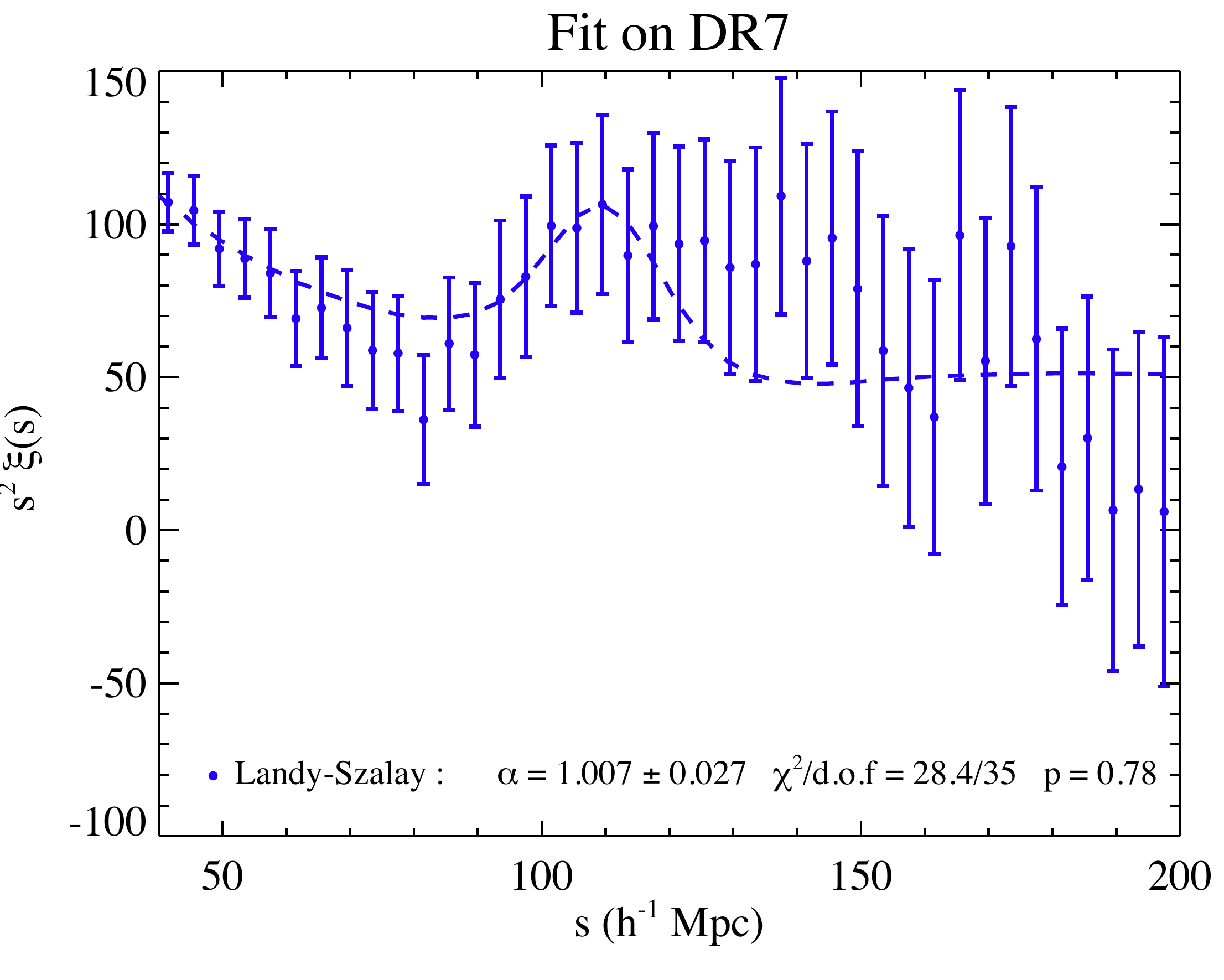}
  \includegraphics[width=0.49\textwidth]{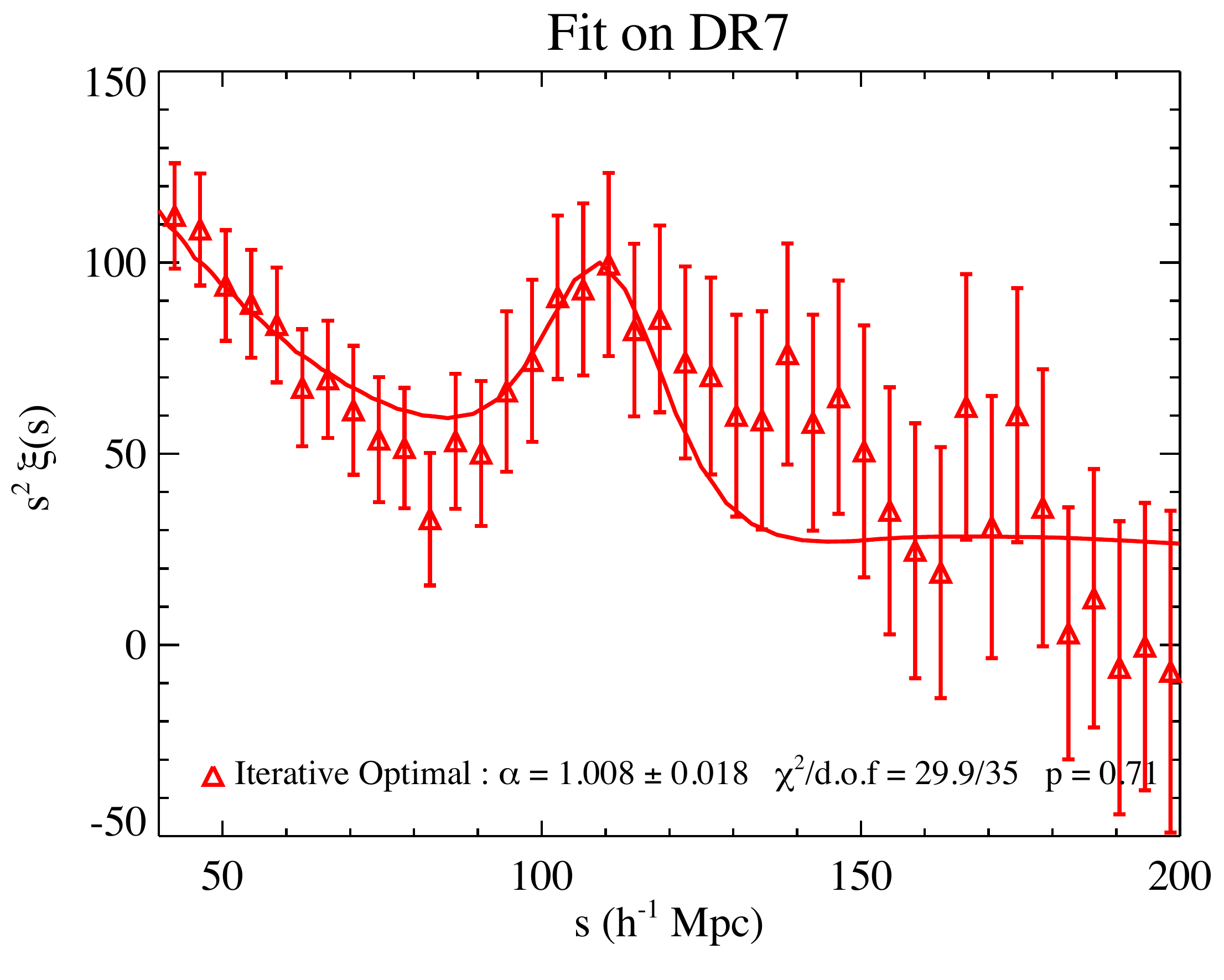}
 	\includegraphics[width=0.49\textwidth]{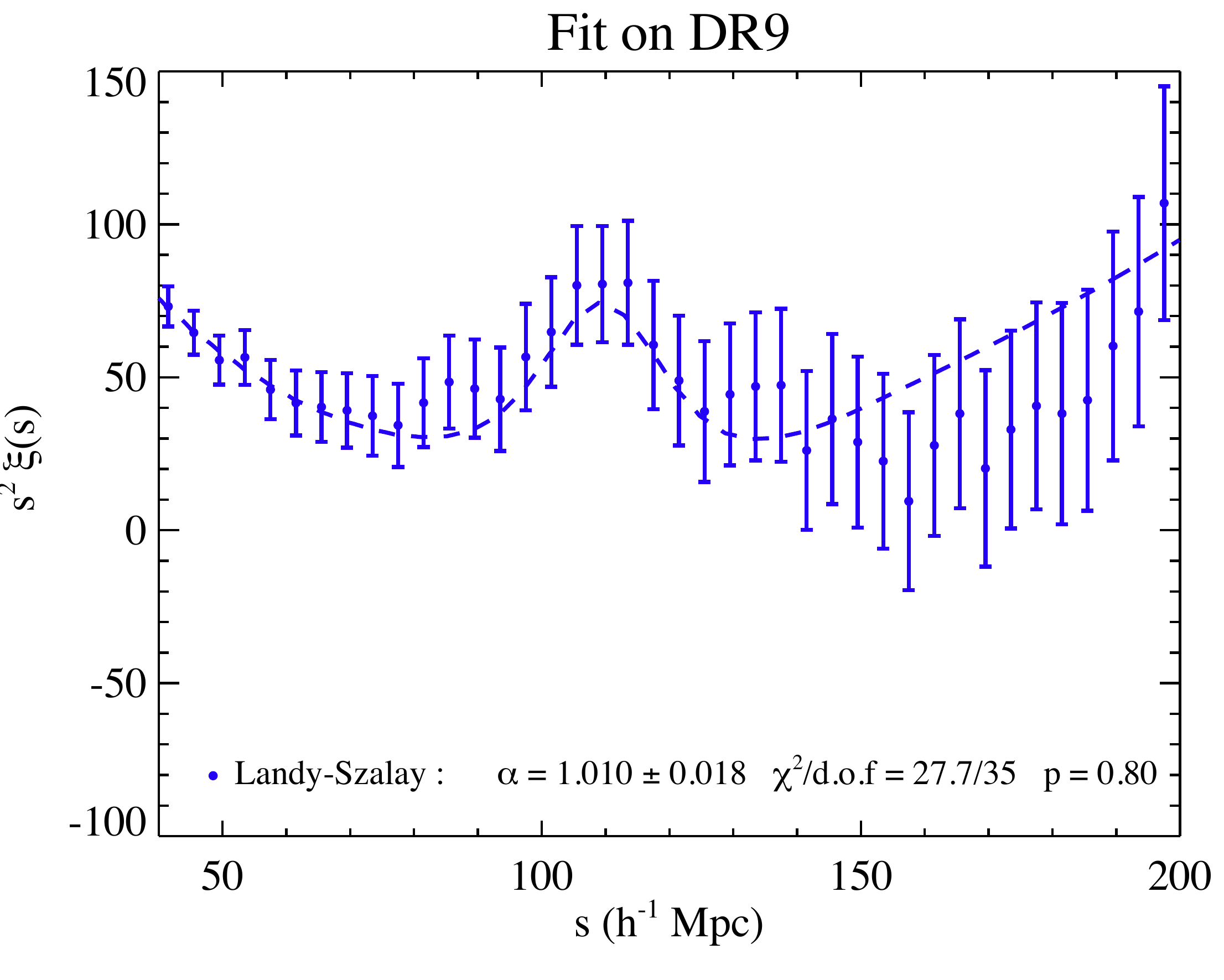}
	\includegraphics[width=0.49\textwidth]{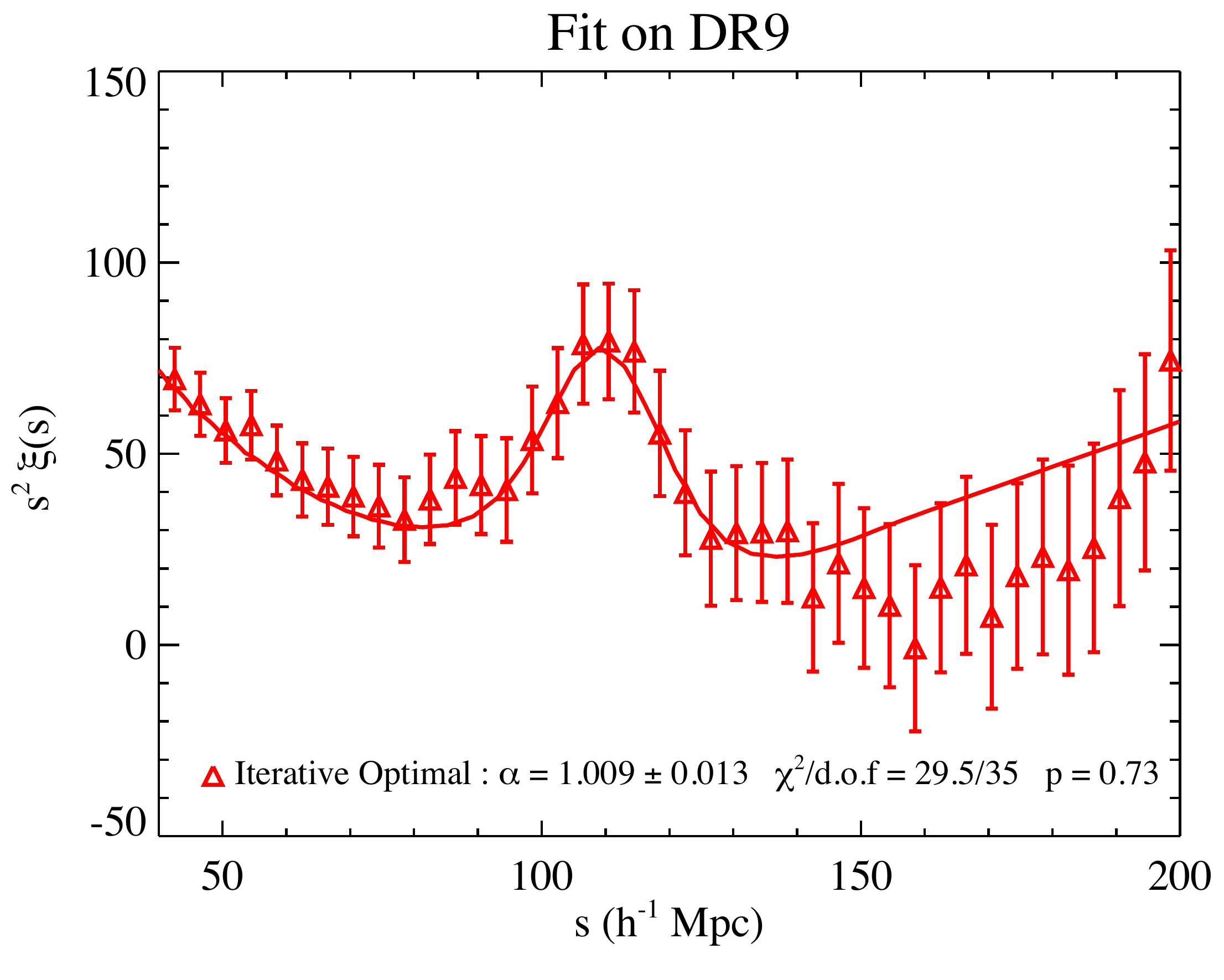}}
   \caption{ Correlation functions obtained for DR7 LRG (top) and DR9 CMASS (bottom) data samples using the Landy-Szalay (left panels) and the iterative optimal estimator (right panels). The corresponding best fit is shown in solid red  for Landy-Szalay and dashed blue for the iterative optimal estimator, and $\alpha_\mathrm{Measured}$ is given in the legend, together with the $\chi^2/$d.o.f. and its probability. The blue points are shifted by 1Mpc/h to improve visibility.
      The covariance matrices used for these fits are based upon the LasDamas (DR7) or  PTHalos (DR9) mock catalogues. 
      (Coloured version of the figure available online)}
      
	\label{fig:dr7_fit}%
\end{figure*}
\subsection{DR9}

Following the same procedure as for the DR7 LRG sample, we computed the spherically averaged two-point correlation function using the Landy-Szalay estimator and the iterative optimal estimator. The results as shown in the bottom panels of  Fig.~\ref{fig:dr7_fit}, we insist again that the plots do not show the correlation between different separation bins. The corresponding values of $\alpha$ are given in Table ~\ref{tab:alphas}.  The $\chi^2$/d.o.f and p-values are shown in the legend of the figure, where we get $\chi^2$/d.o.f = 27.7/35 and $p=0.80$ for Landy-Szalay (left panel) and $\chi^2$/d.o.f = 29.5/35 and $p=0.73$ for the optimal estimator (right panel), indicating a good fit in both cases.

We see clear improvement in the precision of the $\alpha$ measurement compared to the Landy-Szalay one. The values agree, but the iterative estimator gives us a 28\% more accurate result.

As discussed for the DR7 data, $\alpha_\mathrm{Measured}$ and its error for DR9 CMASS data are consistent with the measurements with PTHalos mocks, as can be seen in Figs.~\ref{fig:scatter_alpha} (right) and \ref{fig:scatter_error} (right).

The BOSS DR9 CMASS result [\cite{anderson}], using the correlation function, only is $\alpha~=~1.016~\pm~0.017$ before and $\alpha~=~1.024~\pm~0.016$ after reconstruction; however in the case of DR9, the improvement of 6\% due to reconstruction is much lower that expected with the new iterative estimator. Meanwhile,       this result is consistent with our values with both estimators (Table.~\ref{tab:alphas}) well within 1-$\sigma$.

\begin{table}
\centering
\caption{Values of $\alpha$ found with two different estimators of the correlation function for each sample. }
\label{tab:alphas}
\begin{tabular}{lccc}
\hline
\hline
Sample & Landy-Szalay & It. opt. est. & Gain\\
& $\alpha_\mathrm{LS}$ & $\alpha_\mathrm{opt}$ & \% \\
\hline
Mean LasDamas &	$0.976 \pm 0.035$ & 	$0.979 \pm 0.029$ & 17\\
Mean PTHalos & 	$1.013 \pm 0.039$ &	$1.011 \pm 0.031$ & 21\\
DR7 			& 	$1.004 \pm 0.026$ & 	$1.006 \pm 0.018$ & 31\\
DR9 			& 	$1.010 \pm 0.018$ &	$1.009 \pm 0.013$ & 28\\
\hline
\hline
\end{tabular}
\end{table}

\begin{figure*}
   \centering
   	{\includegraphics[width=0.4\textwidth]{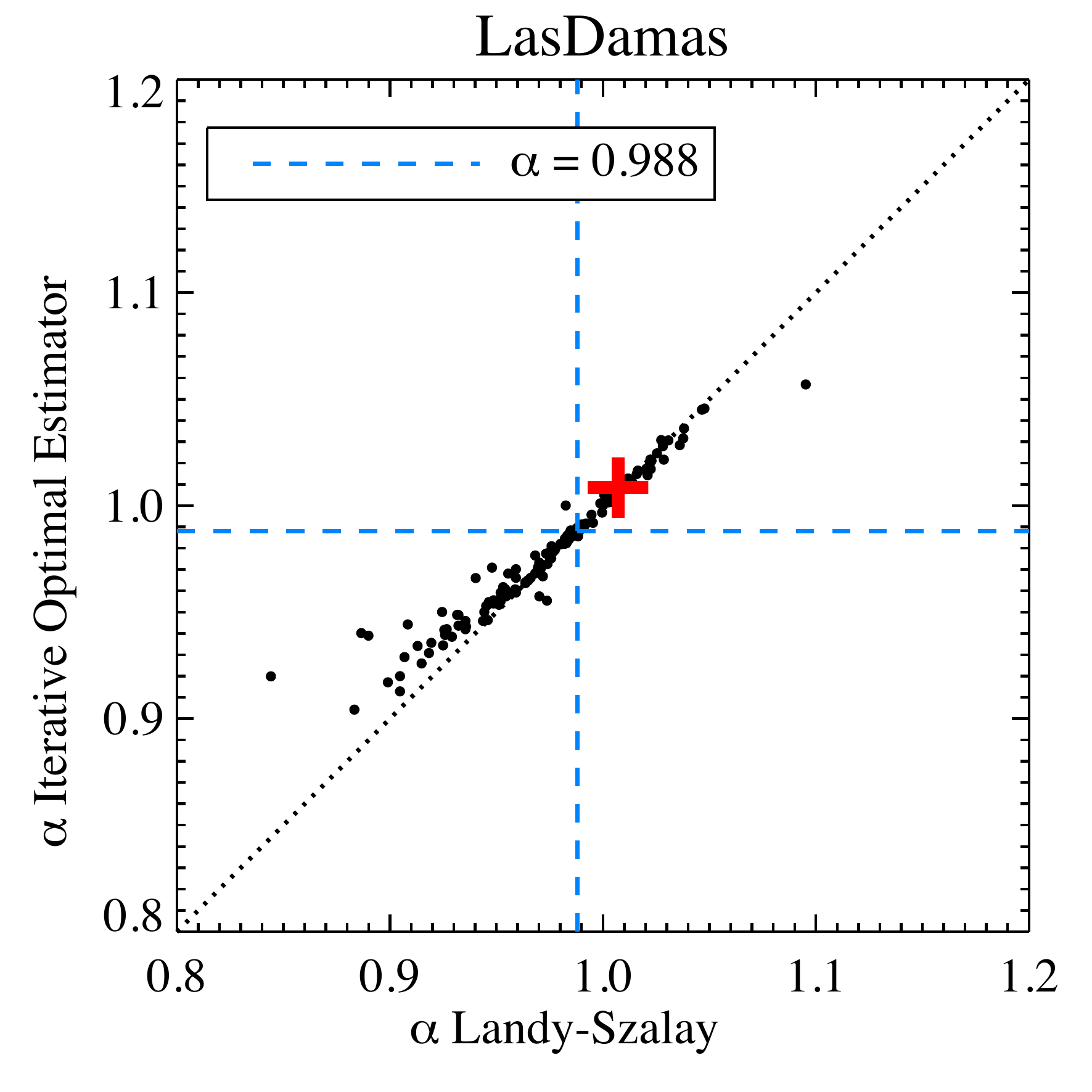}	
  			\includegraphics[width=0.4\textwidth]{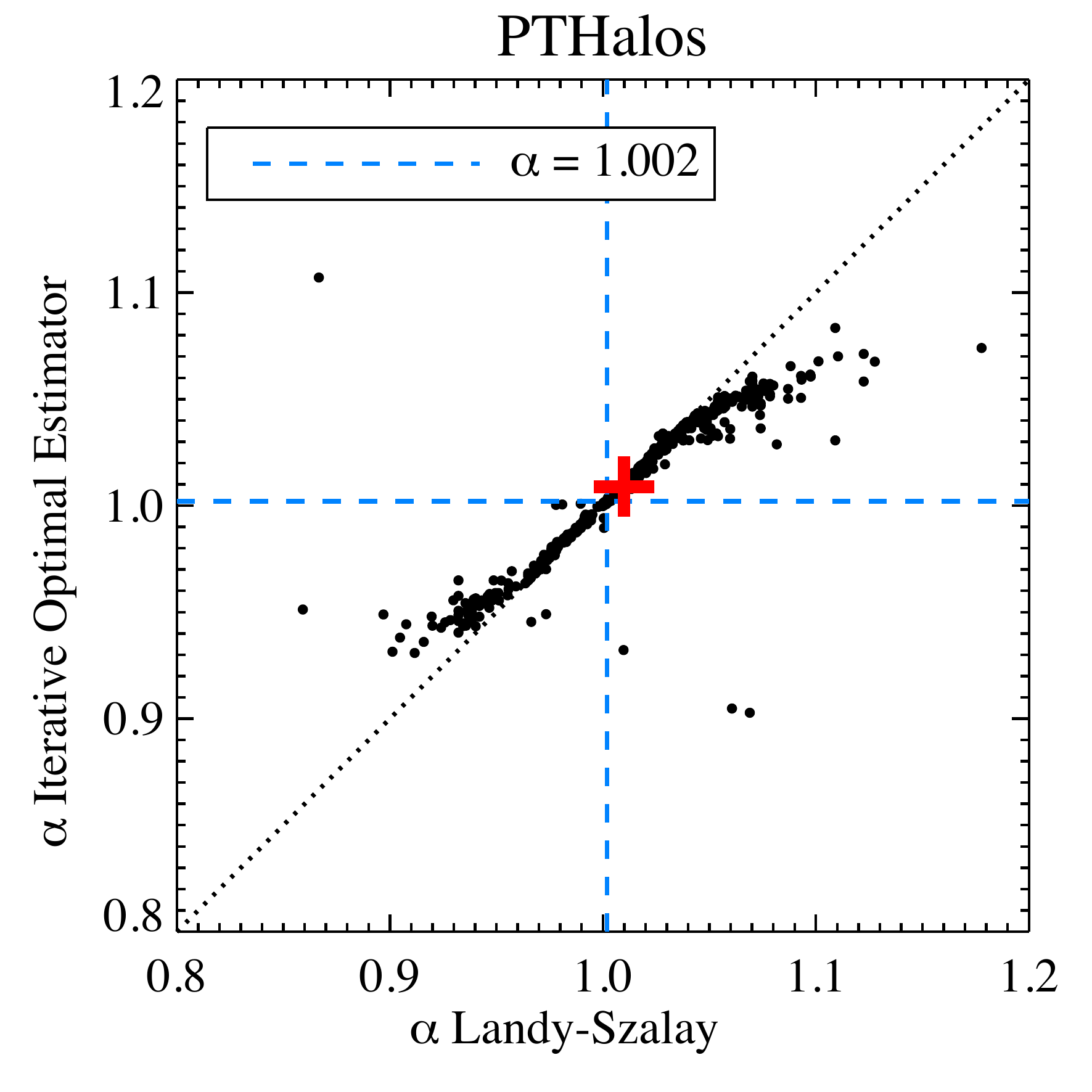}}
   \caption{Comparison of $\alpha_\mathrm{Measured}$ using the Landy-Szalay and the iterative optimal estimators for the mocks (points) and the real data (crosses) for LasDamas and DR7 (left), and PTHalos and DR9 (right). The expected values of $\alpha$ are shown by dashed lines.
   (Coloured version of the figure available online)
   }
              \label{fig:scatter_alpha}%
\end{figure*}

\begin{figure*}
   \centering
         {\includegraphics[width=0.4\textwidth]{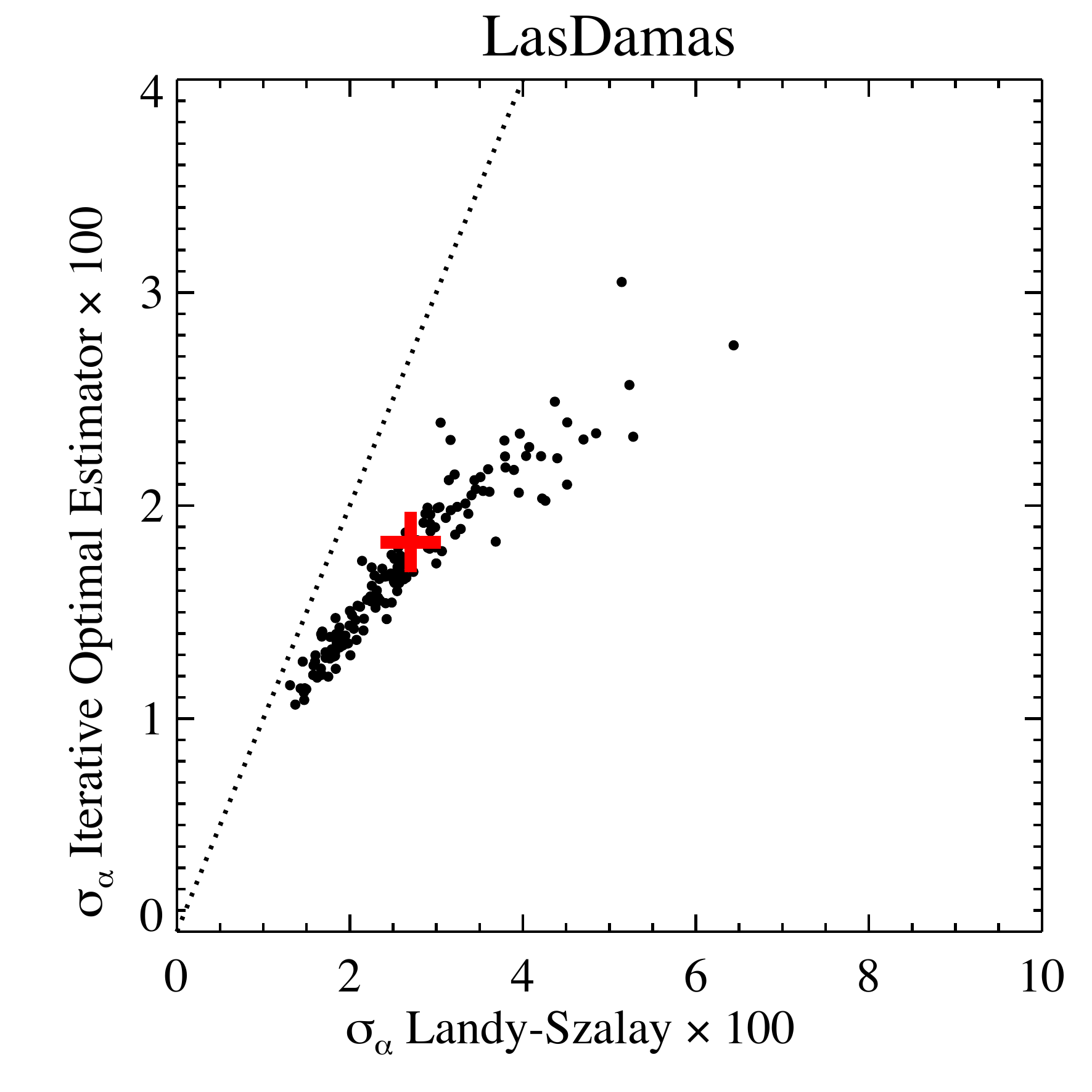}	
  						\includegraphics[width=0.4\textwidth]{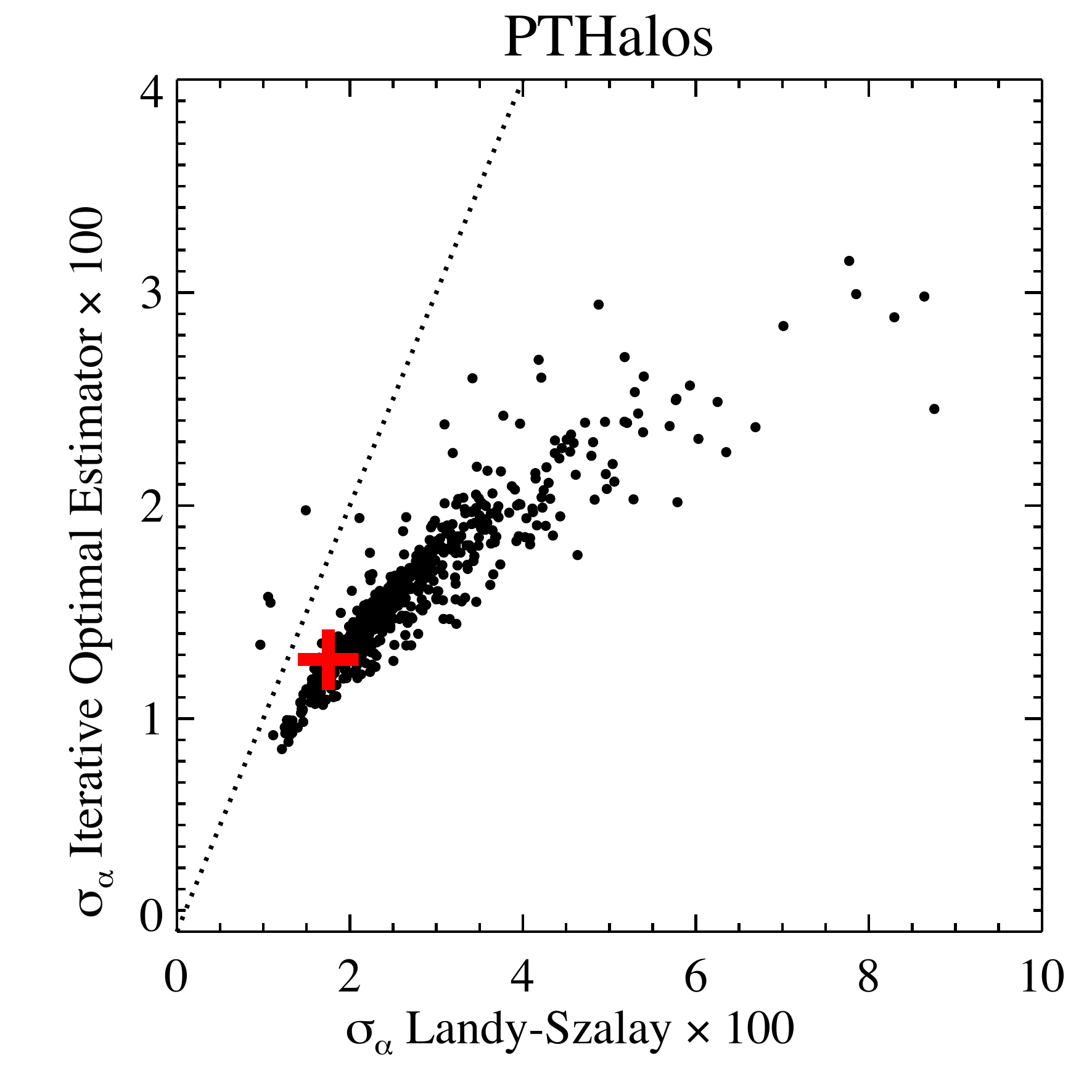}}
   \caption{Comparison of the error on $\alpha_\mathrm{Measured}$ using Landy-Szalay and the iterative optimal estimators for the mocks (points) and the real data sample (crosses). As in Fig.~\ref{fig:scatter_alpha}, the left plot is for LasDamas and DR7 LRG data and the right one for PTHalos and DR9 data. The dotted line corresponds to the same error for the two estimators. 
   (Coloured version of the figure available online). 
   }
              \label{fig:scatter_error}%
\end{figure*}

\section{Cosmological constraints}

The improvement on cosmological parameter constraints using the iterative optimal estimator is illustrated in Fig.~\ref{fig:constraints}.  
These constraints are obtained using a Monte Carlo Markov Chain within an open $\Lambda$CDM cosmology using CMB data alone. The chain\footnote{The MCMC from WMAP7 is available at \texttt{http://lambda.gsfc.nasa.gov/}.} was resampled with our BAO $\alpha$ constraints.
The marginalized constraints on $\Omega_m$ and $\Omega_\Lambda$ are given in Table.~\ref{tab:const}. 

The overall gain on the cosmological parameters is between 13\% and 22\% (except for $\Omega_\Lambda$ for DR9). With the iterative optimal estimator applied to DR7 data, the accuracy on $\Omega_m$ and $\Omega_\Lambda$ is comparable to what is measured with the Landy-Szalay estimator applied to the DR9 sample, even though the DR9 has a density that is three times higher and has twice the volume of DR7.

\begin{table}[h]
\caption{Improvement in cosmological parameters with the iterative optimal estimator.}
\label{tab:const}
\begin{center}
\begin{tabular}{l|ll|ll}
\hline
\hline
WMAP7+ 			& $\Omega_m$ 		&Gain  	&$\Omega_\Lambda$ 	&Gain\\
\hline
DR7 (LS) 		& $0.276 \pm 0.018$	&   - 	&$0.727 \pm 0.017$ 	& -  	\\
DR7 (It. Opt.) 	& $0.274 \pm 0.014$ 	& 22\% 	&$0.729 \pm 0.014$ 	& 17\%	\\
DR9 (LS) 		& $0.278 \pm 0.015$ 	&	- 	& $0.725 \pm 0.015$ 	& -		\\
DR9 (It. Opt.) 	& $0.278 \pm 0.013$ 	&  13\% 	& $0.725 \pm 0.015$ 	& 0\%	\\
\hline
\hline
\end{tabular}\end{center}
\end{table}

\section{Conclusions}

We have designed a new two-point correlation function estimator, which is  a linear combination of all possible ratios (up to the second order) of pair counts between data and random samples. The linear combination can be optimized to minimize the variance of the correlation function for a given geometry. We developed an iterative procedure to make this new estimator independent of the cosmology of the simulated data used in its optimization.  We have shown on lognormal, second-order perturbation theory and N-body simulations that the decrease in size of the correlation function error bars is around 25\%, relative to the well known Landy-Szalay estimator. The improvement is not mitigated by extra correlations in the covariance matrix of the two-point correlation function.

This result is not contradictory with the fact that the Landy-Szalay estimator was shown to be of minimal variance, since this is only true for a vanishing correlation function and simple geometry. Current galaxy surveys do measure a non-zero correlation function even on large scales, and they have quite complex  geometry.

Our method can be easily applied to any dataset  and requires few extra mock catalogues compared to the standard Landy-Szalay-based analysis. 
Extra mock catalogues with different cosmologies are required to evaluate the iterative optimal estimator that we have shown is unbiased. In our implementation we used lognormal simulations to produce these %
extra mock catalogues, since the production of  lognormal mocks is straightforward and fast, but our method could be adapted to any kinds of simulations. 
The method requires modest extra CPU time (in our implementation, just a few days on a single desktop machine).

Finally, we have applied our method to SDSS DR7 and DR9 data, achieving an improvement of 10-15\% on the value of the cosmological parameters $\Omega_m$ and $\Omega_\Lambda$. We achieve a similar accuracy with our estimator on the DR7 sample as with the Landy-Szalay estimator on the DR9 sample.

For future developments, we would use principal component analysis to identify the combination of ratios that contributes most to minimize the correlation function variance. The optimization could then be limited to the most relevant combinations. This method can be easily extended to the study of the anisotropic correlation function. The coefficients would be optimized to simultaneously minimize the variance of the monopole and quadrupole. That approach would produce better constraints on redshift space distortions [\cite{kaiser1987}] physical parameters and on the Alcock-Paczynski test [\cite{APtest1979}].

The optimized iterative estimator could be easily applied to marked correlation functions as well  [e.g., \cite{Skibba2006}; \cite{Martinez2010}]. 

\begin{figure}
   \centering
   \resizebox{\hsize}{!}{\includegraphics{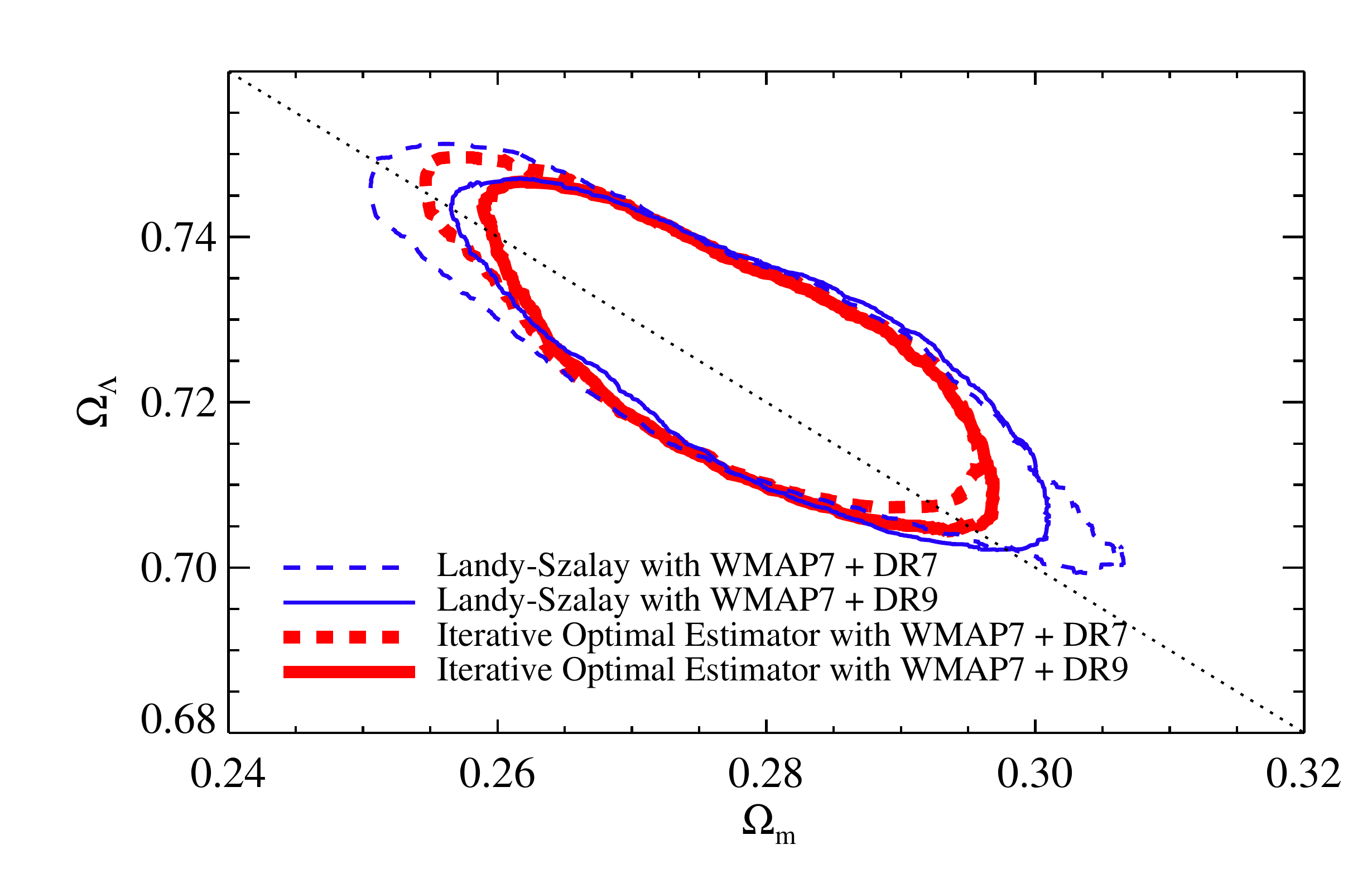}}
   \caption{The 68\% joint constraints in the $(\Omega_m,\Omega_\Lambda)$ plane for an open $\Lambda$CDM cosmology combining CMB (WMAP 7 years~[\cite{WMAP}]) and either DR7 (dashed lines) or DR9 (solid lines) SDSS BAO data, with either the Landy-Szalay estimator (blue thin lines) or the iterative optimal estimator (red thick lines). 
   (Coloured version of the figure available online).}
   
              \label{fig:constraints}%
\end{figure}

\begin{acknowledgements}
      We would like to thank the SDSS-III collaboration for such wonderful data. We thank N. Padmanabhan and J.K. Parejko for making their kd-tree code available. 
      
      We used the``gamma" release LRG  galaxy mock catalogues produced by the
LasDamas projecta and we thank the LasDamas collaboration for providing us with this data. 
     
     We would like to thank R. Skibba, Chia-Hsun Chuang, Lado Samushia, and Graziano Rossi for helpful suggestions and comments.
     
This project was supported by the Agence Nationale de la Recherche under contract ANR-08-BLAN-0222.

Funding for SDSS-III has been provided by the Alfred P. Sloan Foundation, the Participating Institutions, the National Science Foundation, and the U.S. Department of Energy Office of Science. The SDSS-III web site is http://www.sdss3.org/.

SDSS-III is managed by the Astrophysical Research Consortium for the Participating Institutions of the SDSS-III Collaboration including the University of Arizona, the Brazilian Participation Group, Brookhaven National Laboratory, University of Cambridge, Carnegie Mellon University, University of Florida, the French Participation Group, the German Participation Group, Harvard University, the Instituto de Astrofisica de Canarias, the Michigan State/Notre Dame/JINA Participation Group, Johns Hopkins University, Lawrence Berkeley National Laboratory, Max Planck Institute for Astrophysics, Max Planck Institute for Extraterrestrial Physics, New Mexico State University, New York University, Ohio State University, Pennsylvania State University, University of Portsmouth, Princeton University, the Spanish Participation Group, University of Tokyo, University of Utah, Vanderbilt University, University of Virginia, University of Washington, and Yale University.

\end{acknowledgements}
\bibliographystyle{aa} 
\bibliography{optimal} 
\end{document}